\documentclass[11pt]{article}
\usepackage{amsmath,amssymb,color,graphics,epsfig}
\usepackage[utf8x]{inputenc}
\usepackage[unicode=false,
bookmarks=true,bookmarksnumbered=false,bookmarksopen=false,
breaklinks=false,pdfborder={0 0 1},backref=false,colorlinks=false]
{hyperref}

\usepackage[shadow,textwidth=2.6cm]{todonotes}
\usepackage{ifthen}
\reversemarginpar
\newcounter{todocounter}
\definecolor{aqua}{rgb}{0.0, 1.0, 1.0}
\colorlet{ypcolor}{green!40!white}

\newcommand{\ypinline}[2][]{
	
	\ifthenelse { \equal {#1} {} }
	
	{ \def\temp {#2} }  
	
	{ \def\temp {#1} }   
	
	\refstepcounter{todocounter}\todo[color=ypcolor,inline,caption={\textbf{\thetodocounter. YP} \temp}]{\textbf{\thetodocounter. YP:} #2}{}}

\colorlet{hpjcolor}{blue!40!white}

\newcommand{\hpjinline}[2][]{
	
	\ifthenelse { \equal {#1} {} }
	
	{ \def\temp {#2} }  
	
	{ \def\temp {#1} }   
	
	\refstepcounter{todocounter}\todo[color=hpjcolor,inline,caption={\textbf{\thetodocounter. HPJ} \temp}]{\textbf{\thetodocounter. HPJ:} #2}{}}

\usepackage{amsfonts}

\newcommand{\hoch}[1]{$\, ^{#1}$}

\newcommand{\be}{\begin{equation}}
\newcommand{\ee}{\end{equation}}
\newcommand{\bea}{\setlength\arraycolsep{2pt} \begin{eqnarray}}
\newcommand{\eea}{\end{eqnarray}}
\newcommand{\nn}{\nonumber}

\def\ft#1#2{{\textstyle{\frac{\scriptstyle #1}{\scriptstyle #2} } }}

\def\0{{\sst{(0)}}}
\def\1{{\sst{(1)}}}
\def\2{{\sst{(2)}}}
\def\3{{\sst{(3)}}}
\def\4{{\sst{(4)}}}
\def\5{{\sst{(5)}}}
\def\6{{\sst{(6)}}}
\def\7{{\sst{(7)}}}
\def\8{{\sst{(8)}}}
\def\sst#1{{\scriptscriptstyle #1}}

\def\a{\alpha}
\def\b{\beta}

\def\l{\lambda}
\def\L{\Lambda}
\def\m{\mu}
\def\n{\nu}
\def\r{\rho}
\def\s{\sigma}

\def\t{\tau}

\def\tr{\rm tr}

\begin{document}
\begin{center}
{\Large {\bf Force-free higher derivative Einstein-Maxwell theory and multi-centered black holes}}

\vspace{20pt}

{\large Peng-Ju Hu\hoch{1} and Yi Pang\hoch{1,2}}

\vspace{10pt}

{\it \hoch{1} Center for Joint Quantum Studies and Department of Physics,\\
School of Science, Tianjin University, Tianjin 300350, China }
\bigskip

{\it \hoch{2}Peng Huanwu Center for Fundamental Theory,\\
Hefei, Anhui 230026, China}

\vspace{40pt}

\underline{ABSTRACT}
\end{center}
\par

We investigate which 4-derivative extensions of Einstein-Maxwell theory admit multi-extremal black hole solutions 
with gravitational force balanced by Coulomb force. We obtain a set of constraints on the 4-derivative couplings by exploring various probe limits in multi-black hole systems. It turns out that these constraints are tighter than those needed to protect the mass-charge ratio of extremal black holes from higher derivative corrections. In fact, they are so strong that the Majumdar-Papapetrou multi-black solutions are unmodified by the force-free combinations of the 4-derivative couplings. Explicit examples of such 4-derivative couplings are given in 4-and 5-spacetime dimensions. Interestingly these include curvature-squared supergravity actions and the quasi-topological $F^4$ term.

\vfill{\footnotesize   \ \ \ pengjuhu@tju.edu.cn\ \ \ pangyi1@tju.edu.cn}

\thispagestyle{empty}
\pagebreak

\setcounter{tocdepth}{3} 
\tableofcontents 
\newpage
\section{Introduction}
In Newtonian mechanics, a collection of charged particles, each having mass equal to its charge (in proper units), can stay still, reaching a state of mechanical equilibrium. Remarkably, the balance between the attractive gravitational force and the repulsive Coulomb force still holds in general relativity where this system is described by the Majumdar-Papapetrou (MP) multi-black hole solution \cite{Majumdar:1947eu, Papaetrou:1947ib} in Einstein-Maxwell (EM) theory without a cosmological constant. The MP solution is free of any conical singularity, since each black hole is an extremal Reissner-Nordstr\"{o}m black hole \cite{Hartle:1972ya}. Thus the emsemble of multi-black holes is also in  thermal equilibrium. When a positive cosmological constant is added, the MP solution is generalized to the Kastor-Traschen solution \cite{Kastor:1992nn}.  Whether the analog of MP solution exists in asymptotically AdS space remains an open question while some attempts have been made in recent years \cite{Cai:2022qac, Monten:2021som, Chimento:2013pka, Anninos:2013mfa}.

In higher derivative extensions of Einstein-Maxwell theory, little is known about the existence of multi-extremal black hole solutions. The goal of this paper is to figure out what kind of higher derivative couplings can accommodate multi-centered extremal black hole solutions whose total energy is independent of the position of each black hole. Such a system of multi-black holes reaches internal mechanical and thermal equilibrium  without the aid of external forces, unlike the multi-black holes  supported by strut \cite{Costa:2000kf}, acceleration \cite{Gregory:2020mmi} or cosmic expansion \cite{Dias:2023rde}. Naively, one may think that the higher derivative couplings only modify the solution with no effects on its existence. However, lessons from Weak Gravity Conjecture suggest that 
higher derivative interactions do in general break the balance between gravity and Coulomb force \cite{Kats:2006xp}. Hence existence of multi-extremal black hole solutions will require higher derivative couplings satisfy certain constraints, possibly carving out the boundary between the landscape and the swampland. As we will see, these constraints are stronger than those needed to protect the mass-charge ratio of extremal black holes from higher derivative corrections.

In this paper,  we focus on 4-derivative extensions of Einstein-Maxwell theory composed of Riemann tensor and U(1) field strength. The brute-force way would be to solve the 4-derivative corrections to the multi-black hole solutions and impose certain smoothness conditions. 
This approach is not feasible due to the lack of symmetry for multi-black hole solutions. Instead, we follow a more physical path by first considering a 2-black hole system where one of the black hole is much lighter than the other. In this case, the lighter
black hole is treated as a charged massive particle probing the background sourced by the
heavier one.  The force felt by the test particle must vanish as a result of balance between gravity and Coulomb force. The simple probe limit in the 2-black hole system already yields two constraints on the 4-derivative couplings, of which, one combination implies that the two black holes satisfy the same extreme mass-charge relation as in the 2-derivative theory. Beyond the probe limit, backreactions from the test particle to the background solution are considered up to first order in the mass of the test particle. The force-free condition derived using the backreacted solution leads to surprisingly strong constraints on the 4-derivative couplings such that the multi-centered MP solutions are not modified by the force-free combinations of 4-derivative couplings.

The procedure above is carried out explicitly in 4-and 5-spacetime dimensions. In these two dimensions, the Einstein-Maxwell theory is closely related to the bosonic action of $N=2$ supergravity \footnote{In $D=4$, Einstein-Maxwell action is the full bosonic action of $N=2$ supergravity. In $D=5$, the bosonic action contains also the Chern-Simons term which has no effect to the purely electric MP solutions.} with the MP solution preserving half of supersymmetry. Thus  it is conceivable that the force-free 4-derivative terms may correspond to 4-derivative supergravity invariants. Our results show that in $D$=4 and 5, the force-free 4-derivative couplings contains not only the  4-derivative supergravity invariants (parity even terms in $D=5$) but also one more structure which we identified as the $F^4$ term in quasi-topological electromagnetism.

This paper is organized as follows. In section 2, we review the force-free property of extremal black holes in Einstein-Maxwell theory.  In section 3,  we  carry out the procedure outlined above in 4-and 5-dimensions, deriving the 4-derivative couplings that enjoy the force-free property and can accommodate multi-extremal black hole solutions.  We conclude with discussions in section 4.

\section{Static Charged multi-black holes in Einstein-Maxwell theory}

The solutions that will be utilized here are those of the $D$-dimensional Einstein-Maxwell theory
\begin{equation}
	S_{\rm EM}=\frac{1}{2\kappa^{2}}\int d^{D}x\sqrt{-g}(R-\frac{1}{4g^{2}}F_{\mu\nu}F^{\mu\nu})\ .\label{2PartialTheory}
\end{equation}
The widely studied single static charged black hole solution is of the form
\bea
	ds^{2} & =&-f(r)dt^{2}+\frac{dr^{2}}{f(r)}+r^{2}d\Omega_{D-2}^{2}\ ,
\nn\\
	f(r) & =&g(r)=1-\frac{m}{2r^{D-3}}+\frac{q^{2}}{4r^{2(D-3)}}\ ,
\nn\\
 A_{(1)}&=&-a_{D}\frac{g q}{r^{D-3}}dt\ ,\quad a_{D}=\sqrt{\frac{D-2}{2(D-3)}}\ ,
\label{DBHq}
\eea
where $d\Omega_{D-2}^{2}$ is the line element on a unit $(D-2)$-sphere. To simplify the discussion, from now on, we set U(1) coupling $g=1$. Parameters $m$ and $q$ are related to the ADM mass $M$ and electric charge $Q$ via
\be
m=\frac{2M}{(D-2)A_{D-2}},\quad q=\sqrt{\frac{2}{(D-2)(D-3)}}\frac{Q}{A_{D-2}}\ ,
\ee
where $A_{D-2}$ denotes the area of a unit $(D-2)$-sphere.
In addition, the theory also admits multi-extremal black hole solutions \cite{Majumdar:1947eu,Papaetrou:1947ib,Myers:1986rx}
\be
        ds^{2}  =-H^{-2}dt^{2}+H^{\frac{2}{D-3}}(dx_1^2+\cdots dx_{D-1}^2)\ ,\quad A_t=2a_{D}H^{-1},\quad 
         H =1+\sum_i\frac{m_i}{4|\vec{x}-\vec{x}_i|^{D-3}}\ ,  \label{TwoBH}
\ee
 where $m_{i},\,\vec{x}_{i}$ label the mass and position of each black hole. 
 These black holes have reached the mechanical equilibrium, namely,
the gravitational force is precisely canceled by the Coulomb force. Indeed the $i$-th black hole's mass and charge obeys
\be
M_i=2a_D Q_i\ .
\label{ff}
\ee
In terms of the parameters characterizing a single black hole, it means $m_i=2q_i$. The mass-charge relation is usually called the extremality condition or BPS condition if the Einstein-Maxwell theory is embedded in supergravity. The force-free property is reflected in the fact that the energy of the multi-black hole solutions \eqref{TwoBH} is independent of the position of individual black hole.

In fact, the mass-charge relation  \eqref{ff} needed for the system to be force-free can also be obtained by considering a certain limiting situation without actually knowing the exact multi-black hole solutions. For instance, in the case of two black holes, suppose one of the black hole is much lighter than the other so that at the leading order of the small mass ratio, the lighter black hole can be treated as a charged massive particle probing the background of the heavier black hole. Needless to say, the test particle must be able to stay static is a necessary condition underlying the existence of the 2-centered MP solution. 

The action of a charged massive particle takes the form
\be
        S_0=-m_0\int \sqrt{-g_{\m\n}dx^\m dx^\n}+q_0\int A_\m dx^\m\ ,
\label{s0}
\ee
where the target space coordinates $x^\m$ is pulled back on the world-line of the particle. The relation between $m_0$ and $q_0$ will be determined later. In the background of a single charged black hole, an initially static particle is entitled to remain motionless as long as the potential read off from the action \eqref{s0} is independent of $r$
\be
 V_{0}=m_{0}\sqrt{-g_{tt}}-q_{0} A_t={\bf \text{const.}}\ 
 \label{V0}
\ee
Substituting the black hole solution into \eqref{V0}, we find that 
\begin{equation}
	V_0=m_0-\frac{mm_{0}-4a_{D}q q_{0}}{4r^{D-3}}-\frac{m_{0}(m-2q)(m+2q)}{32r^{2(D-3)}}+\cdots\ ,\label{Vq}
\end{equation}
where ``$\cdots$" are higher order terms in $1/r$. Clearly, for $V_0$ to be $r$-independent, the mass and charge parameters must obey
\begin{equation}
	m=2q\ ,\quad  m_{0}=2a_{D}q_{0}\ .
\end{equation}
Plugging the mass-charge relations above back to $V_0$, we confirm that the potential is indeed constant. 
Thus we see that from the simple consideration of a probe particle in the background of a charged black hole, one indeed recovers the mass-charge relation \eqref{ff}. 

Beyond the probe limit, one must include back-reaction of the probe particle to the background solution.  This eventually leads to a 2-centered MP solution and the force felt by the lighter black hole is not well defined at its own location, since the metric and electric potential diverge right there. Away from  the center of black holes, the force-free property is manifest in the MP solution as $A_t=2a_D\sqrt{-g_{tt}}$. The consideration of the probe limit applies to systems with more than 2 black holes as well. For example, in a 3-black hole system, when one of the black hole is much lighter than the other two, the lightest one can be treated as a test particle probing the 2-black hole solution. For Einstein-Maxwell theory, this again leads to the extreme mass-charge ratio satisfied by the third black hole.

So far, we see that for Einstein-Maxwell theory, the probe limit adopted above reproduces the known mass-charge relation of extremal black holes. In the next section,  we apply the same idea  to study existence of multi-extremal black hole solutions in 4-derivative extensions of Einstein-Maxwell theory. It is shown that besides the mass-charge relation \eqref{ff}, the force-free property also requires the coefficients of higher derivative couplings satisfy certain constraints. In other words, not all the higher derivative couplings allow the existence of multi-extremal black holes. 
\section{Force-free 4-derivative action}
 In this section, we propose a general method of deriving the force-free higher derivative couplings. For technical viability, we  mainly focus on $D=4,5$. Interestingly,  the 4-derivative couplings determined from the force-free condition are strong enough that ``there-would-be" corrections from the 4-derivative couplings to the multi-extremal black hole solutions vanish. This means that the multi-extremal black hole solutions in the 2-derivative Einstein-Maxwell theory persist to be exact solutions of the force-free higher derivative theory. 

\subsection{The set up}

 In this  section, we explain our method of obtaining the higher derivative couplings that enjoy the force-free property. We start from Einstein-Maxwell theory extended by 4-derivative interactions of the form 
\begin{equation}
	S_{{\rm EM}+4\partial}=\frac{1}{2\kappa^{2}}\int d^{D}x\sqrt{-g}(R-\frac{1}{4}F_{\mu\nu}F^{\mu\nu}+ \Delta\mathcal{L})\ ,\label{4dL}
\end{equation}
where for the time being, the 4-derivative interactions are chosen to be parity even
\bea
	\Delta\mathcal{L}&=& c_{1}R^{2}+c_{2}R_{\mu\nu}R^{\mu\nu}+c_{3}R_{\mu\nu\rho\sigma}R^{\mu\nu\rho\sigma}
 \nn \\
	&& + c_{4}RF_{\mu\nu}F^{\mu\nu}+c_{5}R_{\mu\nu}F^{\mu\rho}F^{\nu}{}_{\rho}+c_{6}R_{\mu\nu\rho\sigma}F^{\mu\nu}F^{\rho\sigma}
\nn \\
        &&+c_{7}F_{\mu\nu}F^{\mu\nu}F_{\rho\sigma}F^{\rho\sigma}+c_{8}F_{\mu\nu}F^{\nu\rho}F_{\rho\sigma}F^{\sigma\mu}+c_9 \nabla^\m F_{\m\n}\nabla_\rho F^{\rho\n}\ .
 \label{L4even}
\eea
The coefficients $c_1\cdots c_9$ may arise from integrating out heavy degrees of freedom at UV energy scale $\L$, thus  $c_1\cdots c_9$ are suppressed by $1/\L^2$.  
 From the viewpoint of effective field theory, $c_i$ ought to be small and thus we will solve the field equations up to first order in $c_i$. At this order, the effect of 4-derivative corrections is encoded in an effective energy-momentum tensor and an effective electric current built upon the leading order solution \cite{Campanelli:1994sj}. Specifically, we shall solve field equations below only up to first order in $c_i$ 
\bea
R_{\m\n}-\ft12 g_{\m\n}R&=&\ft12 (F_{\mu}^{\ \rho}F_{\nu\rho}-\frac{1}{4}g_{\mu\nu}F_{\rho\sigma}F^{\rho\sigma})+\ft12 \Delta T_{\m\n}[g^{(0)}_{\l\s}\,,A^{(0)}_\s]\ ,
\nn\\ 
\nabla_{\mu}F^{\mu\nu}&=&\Delta J^\nu[g^{(0)}_{\l\s}\,, A^{(0)}_\s]\ ,
\label{4deom}
\eea
where the effective energy-momentum tensor and the effective electric current are defined by
\be
\Delta T_{\m\n}=\frac{-2}{\sqrt{-g}}\frac{\delta(\sqrt{-g}\Delta{\cal L})}{\delta g^{\mu\nu}}\,,\quad \Delta J^\nu=\frac{\delta(\Delta{\cal L})}{\delta A_{\nu}}\ ,
\ee
whose explicit forms are given in appendix A. 

To find the force-free 4-derivative extensions of Einstein-Maxwell theory, 
we consider the simplified scenario in which one of the black hole is treated as
a test particle in the background of the other heavier black holes. At first sight, one should also extend the minimally coupled particle action to include higher derivative corrections of the same order. Dimensional analysis suggests the action \eqref{s0} can be extended to
\bea
    S'_0&=&-m_0\int \sqrt{-(g_{\m\n}+\a_1 g_{\m\n}R+\a_2 R_{\m\n}+\a_3 g_{\m\n} F^2+\a_4 F_{\m\s}F_{\n}{}^{\s})dx^{\m} dx^{\n}}
\nn\\
        &&+q_0\int (A_\m+\a_5 \nabla^\n F_{\n\m}) dx^\m\ ,
\eea
where coefficients $\a_i$ all have dimensions of $\L^{-2}$ in accord with the coefficients of 4-derivative couplings in the gravity action. As a matter of fact,  these higher derivative corrections
to the test particle action can be absorbed into redefinition of $g_{\m\n}$ and $A_\m$. In terms of the redefined metric and the U(1) gauge field, the test particle action retains the minimally coupled form, while  the gravity action is still parameterized by \eqref{L4even} with certain new coefficients $c'_i$. Therefore, we can choose a frame in which the 4-derivative gravity action takes the most general form \eqref{L4even}, while the test particle action remains minimally coupled.  Notice that although the test particle action is not modified by higher derivative terms, the background black hole solutions do in general receive corrections from the 4-derivative interactions. 

 In our approach, we first consider probe limits in 2-and 3-black hole systems. In the 2-black hole case,  we only need to solve for 4-derivative corrections to a single charged black hole solution which is straightforward. This leads to 2 constraints on the 4-derivative couplings, of which, a combination implies the two black holes satisfy the same extreme mass-charge relation as in the 2-derivative case. In the 3-black hole case,  we need to solve for 4-derivative corrections to 2-centered extremal charged black hole solutions. Obviously,  a direct construction of such solution is very hard. However, we can consider a special situation where there is a hierarchy of masses in the 3-black hole system $m_0\ll m_2\ll m_1$. While the black hole with mass $m_0$ is treated as the probe, 4-derivative corrections to 2-centered extremal black hole solution with $m_2\ll m_1$ can be solved order by order in $m_2$ in the far zone $r\gg m_{1,2}$. We can then place the test particle in the far zone and require it feel no force from the 2-centered extremal black holes. It turns out that utilizing perturbed solutions up to first order in $m_2$, we already obtain constraints on 4-derivative couplings that ensure multi-extremal black hole solutions to exist. 

To calculate 4-derivative corrections to 2-centered extremal black hole solution, we utilize
the axially symmetric ansatz. In cylindrical coordinates, the metric and the U(1) gauge field take the form 
\be
    ds^{2}_D  =-a(\rho,z)dt^{2}+b(\rho,z)dz^{2}+c(\rho,z)(d\rho^{2}+\rho^{2}d\Omega_{D-3}^{2})\ ,\quad
    A_{(1)}  =2a_{D}u(\rho,z)dt\ .
\label{Cylindrical}
\ee
In terms of spherical coordinates defined by
\be
r^{2}=\rho^{2}+z^{2}\ , \quad z=r\cos\theta\ ,
\ee
the ansatz above (\ref{Cylindrical}) is expressed as
\bea
    ds^{2} &=&-a(r,\theta)dt^{2}+\big(b(r,\theta)\cos^{2}\theta+c(r,\theta)\sin^{2}\theta\big)dr^{2}+r^{2}\big(b(r,\theta)\sin^{2}\theta+c(r,\theta)\cos^{2}\theta\big)d\theta^{2}
\nonumber \\
            &+&c(r,\theta)r^{2}\sin^{2}\theta d\Omega_{D-3}^{2}+\big(c(r,\theta)-b(r,\theta)\big)r\sin2\theta d\theta dr\ ,
\nn\\    
         A_{(1)} &=&2a_{D}u(r,\theta)dt\ .
\label{Ansatz}
\eea
Without higher derivative corrections, the 2-centered extremal black hole in spherical coordinates is given by
\bea
    a_0& =& H(r,\theta)^{-2},\quad b_0=c_0=H(r,\theta)^{\frac{2}{D-3}},\quad u_0=H(r,\theta)^{-1}\ ,
\nonumber \\
    H(r,\theta) & =&1+\frac{m_{1}}{4r^{D-3}}+\frac{m_{2}}{4(r^{2}+z_{2}^{2}-2z_{2}r\cos\theta)^{\frac{D-3}{2}}}\ , 
\label{TwoBH2}
\eea
where the two black holes are located at $z=0,\, z=z_2$ on the polar axis respectively. Notice that the radial coordinate in \eqref{TwoBH2} differs from the radial coordinate in the single black hole solution by the transformation
\be
     \widetilde{r} =r H(r)^{\ft {1}{D-3}}\ ,
\ee
where $\widetilde{r}$ denotes the radial coordinate in the single black hole solution \eqref{DBHq}. One can check it by setting $m_2=0$ and the 2-centered black hole reduces to the extremal single charged black hole solution \eqref{DBHq}. 

When $m_2\ll m_1$, we keep the solution up to linear order in $m_{2}$. The results are 
\begin{align}
    a_0 & =f^{-2}\Big(1-\frac{m_{2}}{2f(r^{2}+z_{2}^{2}-2z_{2}r\cos\theta)^{\frac{D-3}{2}}}\Big)+\mathcal{O}(m_{2}^{2})\ ,
\nn\\
    b_0 & =c_0=f^{\frac{2}{D-3}}\Big(1+\frac{1}{2(D-3)}\frac{m_{2}}{f(r^{2}+z_{2}^{2}-2z_{2}r\cos\theta)^{\frac{D-3}{2}}}\Big)+\mathcal{O}(m_{2}^{2})\ ,
\nn\\
    u_0 & =f^{-1}\Big(1-\frac{m_{2}}{4f(r^{2}+z_{2}^{2}-2z_{2}r\cos\theta)^{\frac{D-3}{2}}}\Big)+\mathcal{O}(m_{2}^{2})\ ,\quad  f(r)=1+\frac{m_{1}}{4r^{D-3}}\ .
\end{align}
In the far zone $r\gg m_{1,2},\,z_2$, we can further expand the solution in (axisymmetric) spherical harmonics as \cite{Candlish:2007fh}
\be
 \frac{1}{(r^{2}+z_{2}^{2}-2z_{2}r\cos\theta)^{\frac{D-3}{2}}}  =\sum_{n=0}^{\infty}\frac{z_{2}^{n}}{r^{D-3+n}}Y_{n}(\cos\theta)\ ,
\ee
where $Y_{n}(\cos\theta)=C_{n}^{\frac{D-3}{2}}(\cos\theta)$, and
$C_{n}^{\frac{D-3}{2}}$ is the $n$-th Gegenbauer polynomial whose first two members are 
\begin{equation}
C_{0}^{\frac{D-3}{2}}(\cos\theta)=1,\ C_{1}^{\frac{D-3}{2}}(\cos\theta)=(D-3)\cos\theta\ .
\end{equation}
Turning on 4-derivative corrections, we adopt the ansatz for the modified 2-centered black hole solution as
\be
    a(r,\theta) =a_0+\delta{a}\ ,\quad b(r,\theta)=b_0+\delta{b}\ ,
\quad
    c(r,\theta) =c_0+\delta{c}\ , \quad u(r,\theta)=u_0+\delta{u}\ ,
\ee
where the corrections  $\{\delta{a},\delta{b},\delta{c},\delta{u}\}$ are solved only up to $\mathcal{{O}}(c_i, m_2 c_i)$. In the far zone, we also expand the perturbations in terms of (axisymmetric) spherical harmonics
\bea
&&\delta{a}=\sum_{n=0}^\infty\sum_{i=0}^\infty\frac{\delta{a}_{(i,n)}}{r^{i}}Y_{n}(\cos\theta) \ ,\quad \delta{b}=\sum_{n=0}^\infty\sum_{i=0}^\infty\frac{\delta{b}_{(i,n)}}{r^{i}}Y_{n}(\cos\theta)\ ,
\nn\\
&&\delta{c}=\sum_{n=0}^\infty\sum_{i=0}^\infty\frac{\delta{c}_{(i,n)}}{r^{i}}Y_{n}(\cos\theta)\ ,  \quad \delta{u}=\sum_{n=0}^\infty\sum_{i=0}^\infty\frac{\delta{u}_{(i,n)}}{r^{i}}Y_{n}(\cos\theta)\ .
\eea

In summary, we will substitute the ansatz into the field equations with 4-derivative corrections 
\bea    a(r,\theta)&=&f^{-2}+\sum_{n=0}^{\infty}\Big(-\frac{m_{2}}{2f^{3}}\frac{z_{2}^{n}}{r^{D-3+n}}+\sum_{i=0}\frac{\delta{a}_{(i,n)}}{r^{i}}\Big)Y_{n}(\cos\theta)+\mathcal{O}(m_{2}^{2})+\mathcal{O}(c_{i}^{2})\ ,
\nn\\
        b(r,\theta)&=&f^{\frac{2}{D-3}}+\sum_{n=0}^{\infty}\Big(\frac{m_{2}f^{\frac{5-D}{D-3}}}{2(D-3)}\frac{z_{2}^{n}}{r^{D-3+n}}+\sum_{i=0}\frac{\delta{b}_{(i,n)}}{r^{i}}\Big)Y_{n}(\cos\theta)+\mathcal{O}(m_{2}^{2})+\mathcal{O}(c_{i}^{2})\ ,
\nn\\       
        c(r,\theta)&=&f^{\frac{2}{D-3}}+\sum_{n=0}^{\infty}\Big(\frac{m_{2}f^{\frac{5-D}{D-3}}}{2(D-3)}\frac{z_{2}^{n}}{r^{D-3+n}}+\sum_{i=0}\frac{\delta{c}_{(i,n)}}{r^{i}}\Big)Y_{n}(\cos\theta)+\mathcal{O}(m_{2}^{2})+\mathcal{O}(c_{i}^{2})\ ,
\nn\\        
        u(r,\theta)&=&f^{-1}+\sum_{n=0}^{\infty}\Big(-\frac{m_{2}}{4f^{2}}\frac{z_{2}^{n}}{r^{D-3+n}}+\sum_{i=0}\frac{\delta{u}_{(i,n)}}{r^{i}}\Big)Y_{n}(\cos\theta)+\mathcal{O}(m_{2}^{2})+\mathcal{O}(c_{i}^{2})\ ,
        \label{SHE}
\eea
and solve for the expansion coefficients  $\{\delta{a}_{(i,n)},\delta{b}_{(i,n)},\delta{c}_{(i,n)},\delta{u}_{(i,n)} \}$ up to  $\mathcal{{O}}(c_i, m_2 c_i)$. Then for the test particle in the background of 2-centered black hole solutions, the force-free condition becomes
\begin{equation}
        V_{0}=m_{0}\sqrt{a(r,\theta)}-2q_{0}a_{D}u(r,\theta)={\rm const}+{\cal O}(c_i^2)\ .
\label{V0n0}
\end{equation}
As we will see, the equation above highly constrains the coefficients of 4-derivative couplings.

\subsection{$D=4$}
In this section, we study consequences of requiring the balance between gravity and Coulomb force in multi-black hole systems with 4-derivative corrections \eqref{4dL}. In practice, this is carried out in the probe limit under which one of the black hole is much lighter 
than the rest and is treated as a charged massive particle.  It turns out that the 2-and 3-black hole cases already yield sufficient constraints on 4-derivative couplings such that the MP solution in the 2-derivative Einstein-Maxwell theory remain to be an exact solution without any corrections. Surprisingly, the full set of constraints obtained from the probe limit of a system involving 3 electric black holes can be reproduced by combining constraints derived from the probe limits of two simpler systems. One consists of 2 electric black holes and the other contains 2 magnetic black holes. 
Furthermore, we will also study force-free condition implied by the probe limit of 2 dyonically charged black holes. This yields the strongest constraints compared to all previous cases. Interestingly, the resulting 4-derivative couplings almost reproduce the 4-derivative invariants in ungauged off-shell $N=2$ supergravity \cite{Bergshoeff:1980is,Butter:2013lta, Bobev:2020egg}  module two unconstrained coefficients $c_1,\,c_9$ which do not affect the solution at all in the first order perturbation theory.
    \subsubsection{Probe limit in the 2-black hole system}

In this subsection, we consider probe limit of the 2-electric black hole solution in which one of the black hole is much lighter than the other so that the lighter black hole is treated as a charged massive particle probing the background of the heavier black hole. To study the implication of force free condition, we first find 4-derivative corrections to the single static charged black hole solution by solving \eqref{4deom} up to first order in $c_i$. The results are given by
\begin{equation}
        ds^{2}=  -f(r)dt^{2}+\frac{dr^{2}}{g(r)}+r^{2}(d\theta^{2}+\sin^{2}\theta d\phi^{2})\,, \quad A=A_{t}dt\ ,
\label{D4qEvenMetric}
\end{equation}
where  
\bea
  f(r)&= & 1-\frac{m}{2r}+\frac{q^{2}}{4r^{2}}-\frac{ q^{2}}{40r^{6}}\Big(\big(c_{2}+2(2c_{3}-20c_{4}-4c_{5}+c_{6}+8c_{7}+4c_{8})\big)q^{2}
\nn\\
        &&+20r^{2}\left(c_{2}+4c_{3}-4c_{4}+2c_{6}\right)-5mr\big(c_{2}+2(2c_{3}-6c_{4}-c_{5}+c_{6})\big)\Big)+\mathcal{O}(c_{i}^{2})\ ,\nn\\
        g(r)&= & 1-\frac{m}{2r}+\frac{q^{2}}{4r^{2}}-\frac{ q^{2}}{40r^{6}}\Big(2\left(3c_{2}+12c_{3}+30c_{4}+11c_{5}+16c_{6}+8c_{7}+4c_{8}\right)q^{2}
\nn\\
        && -5mr\big(3c_{2}+2(6c_{3}+14c_{4}+5c_{5}+7c_{6})\big)+40r^{2}(c_{2}+4c_{3}+8c_{4}+3c_{5}+4c_{6})\Big)+\mathcal{O}(c_{i}^{2})\ ,
\nn\\
        A_{t}&= & -\frac{q}{r}-\frac{1}{20r^{5}}\Big(2q^{3}\left(\tfrac{1}{2}c_{2}+2c_{3}+10c_{4}+c_{5}-9c_{6}-16(2c_{7}+c_{8})\right)+20c_{6}mqr\Big)+\mathcal{O}(c_{i}^{2}),
\eea
which agree with \cite{Chen:2020hjm}. Notice that $c_1$ and $c_9$ do not affect the solution because the  effective energy momentum tensor and electric current associated with $R^2$ and $\nabla^\m F_{\m\n}\nabla_\rho F^{\rho\n}$ vanish on the leading order solution.  Corresponding to the modified solution above, the mass-charge relation of the extremal black hole becomes
\be
m_{\rm ext}=2 q-\frac{8 \big(c_2+4 c_3+2c_5+2c_6+16 c_7+8c_8 \big) }{5 q}+{\cal O}(c_i^2)\ .
\label{ext}
\ee
As discussed before, the action of the test particle retains the same form and thus force-free condition requires the potential be position independent
\begin{equation}
        V_{0}=m_{0}\sqrt{f(r)}-q_{0}A_{t}={\rm const}+{\cal O}(c_i^2)\ .
\label{D4V0q}
\end{equation}
Since the perturbed solution is obtained only up to first order in $c_i$, we impose the force-free condition at the same order. 
Substituting the modified solution \eqref{D4qEvenMetric} into  (\ref{D4V0q}), we obtain 
the mass-charge relation $m=2q,\, m_{0}=2q_{0}$ and two constraints on the higher derivative couplings
\begin{equation}
    c_4= \frac{1}{4} \left(c_2+4 c_3-2 c_6\right)\ , \quad c_5= -\frac{c_2}{2}-2 c_3-c_6-8 c_7-4 c_8\ .
\label{D4qConstraints}
\end{equation}
Notice that the second constraint implies that the mass-charge relation for extremal black hole becomes $m_{\rm ext}=2q$, so that the black hole obeying force-free condition is still extremal. We also see that the force-free condition is in fact stronger than merely keeping the extreme mass-charge ratio unaffected by the higher derivative corrections. In the next subsection, we will explore the consequences of force-free condition using the probe limit in a 3-black hole system. 
\subsubsection{Probe limit in the 3-black hole system}
In a 3-black hole system, we assume there is a hierarchy of masses  $m_0\ll m_2\ll m_1$. The black hole of mass $m_0$ is to be treated as a test particle in the background consisting of 2 heavier black holes which already reach mechanical equilibrium between themselves. This is indicated by the results from the probe limit of 2-black hole case. Without higher derivative corrections, the 2-black hole system is described by the 2-centered MP solution.  
As discussed in section 2, to circumvent the difficulty of solving for the full 4-derivative corrections to the 2-centered MP solution, we expand the 2-centered MP solution in powers of $m_2$ keeping only terms up to first order in $m_2$. Furthermore, we perform large $r$ expansion in the far zone $r\gg m_1,m_2,\,z_2$, where the test particle is placed. Corrections to the approximation of the 2-centered MP can then be solved up to order $c_i,\,m_2 c_i$.  The perturbed solutions associated with each (axisymmetric) harmonic function \eqref{SHE} are independent of each other. Using the solution associated with $Y_0(\cos\theta)$ (given in appendix B), we obtain the following constraints from the force-free condition \eqref{V0n0} 
\begin{equation}
    c_{4}=\frac{1}{4}(c_{2}+4c_{3})\ ,\quad c_{5}=-(c_{2}+4c_{3})\ ,\quad c_{6}=0\ ,\quad c_{7}=\frac{1}{16}(c_{2}+4c_{3}-8c_{8})\ ,
\label{D4qConstraints3}
\end{equation}
which are stronger than the one obtained from the probe limit of the 2-black hole solution \eqref{D4qConstraints}.  Substituting the tightened constraints \eqref{D4qConstraints3} into the effective energy-momentum tensor and effective electric current, we find that 
\begin{equation}
        \Delta T_{\m\n}[g^{(0)}\, ,A^{(0)}]=0\ ,\quad \Delta J^\nu[g^{(0)}\,, A^{(0)}]=0\ ,
\label{JT}
\end{equation}
are satisfied not only on a 2-centered MP solution but actually on all the uncorrected multi-centered MP solutions \eqref{TwoBH}. This result implies that the MP solution in the 2-derivative theory are not modified by the 4-derivative interactions fulfilling the force-free requirement. At this stage, there is no need to explore consequences from the probe limit involving more than 3 black holes. 

Plugging \eqref{D4qConstraints3} in \eqref{L4even}, we obtain the 4-derivative couplings that can accommodate multi-extremal black hole solutions 
\be
       \Delta \mathcal{{L}}_{\text{force-free}}=c'_{1}R^{2}+c_2'{\cal L}_{W^2}+c_3'{\cal L}_{GB}+c_4'\left[(\text{Tr}F^{2})^{2}-2\text{Tr}(F^{4})\right]+c'_{5}\nabla^{\mu}F_{\mu\nu}\nabla_{\rho}F^{\rho\nu}\ ,
\label{D4qLeven}
\ee
where we redefined the coefficients to label the independent structures
\bea
c_1'&=&c_1+\ft13(c_2+c_3)\ ,\quad c_{2}'=\frac{1}{2}(c_{2}+4c_{3})\ ,\quad c_{3}'=-\frac{1}{2}(c_{2}+2c_{3})\ ,
\nn\\
c_{4}'&=&\frac{1}{8}(c_{2}+4c_{3}-4c_{8})\ ,\quad c_{5}'=c_{9}-c_{2}-4c_{3}\ .
\eea
${\cal L}_{W^2}$ and ${\cal L}_{GB}$ correspond to the two 4-derivative invariants in $D=4$ off-shell ungauged $N=2$ supergravity \cite{Bergshoeff:1980is,Butter:2013lta, Bobev:2020egg} upon eliminating the auxiliary fields
\begin{align}
        \mathcal{L}_{W^{2}}=&	C_{\mu\nu\rho\sigma}C^{\mu\nu\rho\sigma}+\ft{1}{2}RF_{\mu\nu}F^{\mu\nu}-2R_{\mu\nu}F^{\mu\rho}F^{\nu}{}_{\rho}-\ft{1}{8}F_{\mu\nu}F^{\mu\nu}F_{\rho\sigma}F^{\rho\sigma}
\nn\\
        &+	\ft{1}{2}F_{\mu\nu}F^{\nu\rho}F_{\rho\sigma}F^{\sigma\mu}+2(\nabla^{\mu}F_{\mu\rho})(\nabla_{\nu}F^{\nu\rho})\ ,
\nn\\        
        \mathcal{L}_{GB}=&	R^{2}-4R_{\mu\nu}R^{\mu\nu}+R_{\mu\nu\rho\sigma}R^{\mu\nu\rho\sigma}\ .
\end{align}
Our result is consistent with and beyond the old result \cite{Kallosh:1992gu} stating that the MP solution does not receive quantum corrections in $N=2$ supergravity, since we also obtain one more structure with coefficient $c_4'$ that preserves the MP solution. The same combination of $F^4$ terms also appeared in quasi-topological electromagnetism \cite{Liu:2019rib,Ma:2020xwi}. In fact, one can prove that the particular combination of two $F^4$ term does not affect all 
purely electric or magnetic solution regardless of the symmetry of the solution. 

 \subsubsection{Concerning magnetic black holes}
In the previous subsection, to obtain the sufficient constraints for the 4-derivative couplings to accommodate 
multi-extremal black hole solution, we have to consider probe limit in a 3-black hole system and solve for corrections to the 2-centered MP solution which is technically more involved compared to the probe limit in the a 2-black hole system. Since Einstein-Maxwell theory also admits magnetically charged black holes, one may wonder what would be the consequences of force-free condition from the probe limit of 2 magnetic black holes although this seems to be unrelated to our original pursuit about existence of multi-electrically charged black holes. To our surprise, the force-free condition imposed on the probe limit of 2 magnetic black holes yields 2 more constraints, which combined with the 2 constraints obtained from the probe limit of 2 electric black holes, leads to the full set of constraints derived from the probe limit in the 3-black hole case. 

To study the probe limit of 2 magnetic black holes, we first find the 4-derivative corrections to a single magnetic black hole. The computation is similar to the electric black hole case. The only difference is that the test particle is now coupled to the dual U(1) gauge field and The outcome of the force-free condition is that 
\be
c_6= \frac{1}{6} \left(-c_2-4 c_3-4 c_4-2 c_5\right)\ ,\quad c_7= \frac{1}{4} \left(c_4-2 c_8\right),
\label{D4pConstraints}
\ee
which when combined with \eqref{D4qConstraints} yields precisely the full set of constraints obtained from the probe limit of 3-centered MP solution. 

To complete the study of force-free condition implied by the probe limit in a 2-black hole system, we consider solutions carrying both electric and magnetic charges. Up to first order in $c_i$, the solution takes the form
\bea
        ds^{2} & =& -f(r)dt^{2}+\frac{dr^{2}}{f(r)}+r^{2}d\Omega_{D-2}^{2}\ ,\quad A=A_{t}dt+p\cos\theta d\phi\ ,
\nn \\
        f(r)&= & 1-\frac{m}{2r}+\frac{p^{2}+q^{2}}{4r^{2}}-\frac{1}{40r^{6}}\Big(c_{2}\left(p^{2}+q^{2}\right)\left(5r(4r-m)+p^{2}+q^{2}\right)
\nn \\
        &&+40r^{2}\left(\left(2c_{3}+2c_{4}+c_{5}+c_{6}\right)p^{2}+\left(2c_{3}-2c_{4}+c_{6}\right)q^{2}\right)
\nn\\
    && -10mr\left(\left(2c_{3}+6c_{4}+2c_{5}+c_{6}\right)p^{2}+\left(2c_{3}-6c_{4}-c_{5}+c_{6}\right)q^{2}\right)
\nn \\
    &&      +4\left(2c_{3}+c_{5}+2c_{6}-8c_{7}\right)p^{2}q^{2}+2\left(2c_{3}-20c_{4}-4c_{5}+c_{6}+8c_{7}+4c_{8}\right)q^{4}
\nn \\
    && +2\left(2c_{3}+20c_{4}+6c_{5}+3c_{6}+8c_{7}+4c_{8}\right)p^{4}\Big)+\mathcal{O}(c_{i}^{2})\ ,
\label{D4qpfr}
\nn\\
 A_{t}&= & -\frac{q}{r}-\frac{1}{20r^{5}}\Big(\big(c_{2}+4c_{3}+20c_{4}+2c_{5}-18c_{6}-32\left(2c_{7}+c_{8}\right)\big)q^{3}
\nn \\
    && +20c_{6}mqr+\left(c_{2}+4c_{3}-20c_{4}-8c_{5}-26c_{6}+64c_{7}\right)p^{2}q\Big)+\mathcal{O}(c_{i}^{2})\ ,
\nn\\
    g(r)&=& 1-\frac{m}{2r}+\frac{p^{2}+q^{2}}{4r^{2}}-\frac{1}{40r^{6}}\Big(2(24c_{3}+7c_{5}+14c_{6}-16c_{7})p^{2}q^{2}+6c_{2}(p^{2}+q^{2})^{2}
\nn \\
    &&+ 2(12c_{3}+30c_{4}+11c_{5}+16c_{6}+8c_{7}+4c_{8})q^{4}-4(-6c_{3}+15c_{4}+2c_{5}+c_{6}-4c_{7}-2c_{8})p^{4}
\nn\\
     &&-  5mr\left(\left(3c_{2}+12c_{3}-28c_{4}-4c_{5}-2c_{6}\right)p^{2}+\left(3c_{2}+12c_{3}+28c_{4}+10c_{5}+14c_{6}\right)q^{2}\right)
\nn \\
    &&+  40r^{2}\left((c_{2}+4c_{3}-8c_{4}-c_{5})p^{2}+(c_{2}+4c_{3}+8c_{4}+3c_{5}+4c_{6})q^{2}\right)\Big)+\mathcal{O}(c_{i}^{2})\ ,
\label{D4qpgr}
\nn\\
   \widetilde{A}_{t}&=&-\frac{p}{r}+\frac{p}{20r^{5}}\left(20c_{6}mr-\left(c_{2}+4c_{3}-20c_{4}-8c_{5}+6c_{6}-64c_{7}-32c_{8}\right)p^{2}\right)
\nn\\
	&&-\frac{pq^{2}}{20r^{5}}\big(c_{2}+2\left(2c_{3}+10c_{4}+c_{5}+7c_{6}+32c_{7}\right)\big)\ ,
\label{D4qpEvenMetric}
\eea
where $p,q$ parameterize the electric and magnetic charge respectively. The purely magnetic black hole we discussed before is obtained from the dyonic black hole \eqref{D4qpEvenMetric} by setting $q=0$. For a dyonically charged test particle, the force-free condition becomes
\begin{align}
        V_{0} & =m_{0}\sqrt{f(r)}-q_{0}A_{t}(r)-p_{0}\widetilde{A}_{t}(r)={\rm const}+{\cal O}(c_i^2)\ ,
\label{D4V0qp}
\end{align}
where $\widetilde{A}_{(1)}$ is the dual U(1) gauge field satisfying $d\widetilde{A}_{(1)}=\star\widetilde{M}_{(2)}, \ \widetilde{M}^{\m\n}=-2\partial\mathcal{L}/\partial F_{\mu\nu}$, with
$\star$ being the Hodge star operator. The dyonic charges should obey \cite{Schwinger:1966nj,Zwanziger:1968rs}
\be
qp_0-pq_0=0\ ,
\ee
so that the electromagnetic field does not carry angular momentum and the solution can be static.

Substituting the modified solution to \eqref{D4V0qp}, we obtain the force-free condition on the mass charge relation
\begin{equation}
        m=2\sqrt{p^{2}+q^{2}},\quad m_{0}=2\sqrt{p_0^{2}+q_0^{2}}\ ,
\label{D4qpConstraints1}
\end{equation}
as well as five constraints on the 4-derivative couplings
\begin{equation}
      c_{4}=\frac{1}{4}\left(c_{2}+4c_{3}\right),\ c_{5}=-\left(c_{2}+4c_{3}\right),\ c_{6}=0,\ c_{7}=-\frac{1}{16}(c_{2}+4c_{3}),\ c_{8}=\frac{1}{4}(c_{2}+4c_{3})\ .
\label{D4qpConstraints2}
\end{equation}
Substituting the constraints above to the general 4-derivative action, we find that the result can be recast into a linear combination of $R^2$, ${\cal L}_{W^2}$, ${\cal L}_{GB}$ and $(\nabla^{\mu}F_{\mu\rho})(\nabla_{\nu}F^{\nu\rho})$ without the quasi-topological electromagnetism term
$(\text{Tr}F^{2})^{2}-2\text{Tr}(F^{4})$. We also noticed that the dyonic black hole solution is not modified by the force-free combination of 4-derivative obtained above. While it is easy to understand why $R^2,\, {\cal L}_{GB}$ and $(\nabla^{\mu}F_{\mu\rho})(\nabla_{\nu}F^{\nu\rho})$ do not affect the solution, it is not obvious that the ${\cal L}_{W^2}$ does not modify the solution neither. Hence we find that two 4-derivative invariants in $D=4$ $N=2$ supergravity not only preserve the supersymmetric MP solution, but also give no corrections to the non-extremal dyonic black holes.
The $F$-dependent terms in ${\cal L}_{W^2}$ (apart from the $(\nabla^{\mu}F_{\mu\rho})(\nabla_{\nu}F^{\nu\rho})$ term) can be expressed in terms of the energy momentum tensor of the U(1) gauge field implying its field equations are thus invariant under electromagnetic duality \cite{Cano:2021tfs}. 
Therefore Einstein-Maxwell theory extended by ${\cal L}_{W^2}$ in fact admits dyonically charged multi-centered extremal black hole solutions obtained via an electromagnetic duality rotation of the purely electric ones.

In $D=4$, one can also consider parity odd 4-derivative couplings 
\bea
\Delta\mathcal{L}_{odd}&=& d_{4}RF_{\mu\nu}\widetilde{F}^{\mu\nu}+d_{5}R_{\mu\nu}F^{\mu\rho}\widetilde{F}_{\ \rho}^{\nu}+d_{6}R_{\mu\nu\rho\sigma}F^{\mu\nu}\widetilde{F}^{\rho\sigma}
\nn \\
&&	+ d_{7}\widetilde{F}_{\mu\nu}F^{\mu\nu}F^{\rho\sigma}F_{\rho\sigma}+d_{8}\widetilde{F}_{\mu\nu}F^{\nu\rho}F_{\rho\sigma}F^{\sigma\mu}\ ,
\eea
where $\widetilde{F}_{\m\n}=\ft12\epsilon_{\m\n\r\l}F^{\r\l}$. The parity odd terms have effects only on the dyonic black holes. Using the probe limit in a 2-dyonic black hole system, we find that the force-free condition selects the special combination
\be
d_5=-4d_4\ , \quad d_6=0\ ,\quad d_8=-4d_7\ .
\ee
Upon using the identity 
\be
g_{\m\n}F_{\r\l}\widetilde F^{\r\l}-4F_{\m\r}\widetilde F_{\n}{}^{\r}=0
\ee
one finds that all the parity odd force-free combinations vanish identically. 
\subsection{$D=5$}
In this subsection,  we continue our pursuit of force-free higher derivative extensions of Einstein-Maxwell in $D=5$.  This time the force-free condition obtained from the probe limit in the 3-electric black hole system leads to five constraints on the 4-derivative couplings, which is one more than those derived in $D=4$ case. This is understandable since in $D=5$, the Gauss-Bonnet term begins to affect solution and thus its coefficient will be constrained by the force-free condition. Another difference between $D=4$ and $D=5$ is that at the 2-derivative level, one can add the Chern-Simons term in the action
\begin{equation}
	{\cal L}_{\rm CS}=\m_{\rm CS}\, A\wedge F\wedge F\ ,
\label{5dL}
\end{equation}
which has no effects on the purely electric black hole solutions.  The analysis is parallel to those carried out in $D=4$. 
    \subsubsection{Probe limit in the 2-black hole system}
    Analogous to the $D=4$, we first consider probe limit of the 2-black hole system in which one of the black hole is much lighter than the other and is treated as a test particle. To study the implication of force free condition in $D=5$, we first find 4-derivative corrections to the single static charged black hole solution by solving \eqref{4deom} up to first order in $c_i$. The results are given by
\begin{align}
        ds^{2}= & -f(r)dt^{2}+\frac{dr^{2}}{g(r)}+r^{2}d\Omega_{3}^{2}\ ,\quad A=A_{t}dt\ ,
\label{ds5q}
\end{align}
where
\bea
        f(r)&= & 1-\frac{m}{2r^{2}}+\frac{q^{2}}{4r^{4}}+\frac{1}{24r^{10}}\Big(24(4c_{4}-c_{5}-4c_{6})q^{2}r^{4}-18(2c_{7}+c_{8})q^{4}
\nn\\
        &&+  q^{2}\left(-12(8c_{4}+c_{5}-2c_{6})mr^{2}+3(23c_{4}+4c_{5}-2c_{6})q^{2}+8(2c_{1}-5c_{2}-22c_{3})r^{4}\right)
\nn\\
        &&+ \frac{1}{4}\left(48c_{3}m^{2}r^{4}+16(-4c_{1}+c_{2}+5c_{3})mq^{2}r^{2}+(47c_{1}+13c_{2}+17c_{3})q^{4}\right)\Big)+{\cal O}(c_i^2)\ ,
\\
        g(r)&= & 1-\frac{m}{2r^{2}}+\frac{q^{2}}{4r^{4}}+\frac{1}{24r^{10}}\Big(-18(2c_{7}+c_{8})q^{4}-72(8c_{4}+3c_{5}+4c_{6})q^{2}r^{4}
\nn\\
        &&-  3q^{2}\left(-4(20c_{4}+7c_{5}+10c_{6})mr^{2}+3(11c_{4}+4c_{5}+6c_{6})q^{2}+8(4c_{1}+5c_{2}+16c_{3})r^{4}\right)
\nn\\
        &&+ \frac{1}{4}\left(48c_{3}m^{2}r^{4}+16(10c_{1}+11c_{2}+31c_{3})mq^{2}r^{2}-(65c_{1}+67c_{2}+191c_{3})q^{4}\right)\Big)+{\cal O}(c^2_i)\ .
\nn\\
        A_{t}&= & -\frac{\sqrt{3}q}{2r^{2}}+\frac{q^{3}\left(-\frac{1}{4}(7c_{1}+5c_{2}+13c_{3})-(9c_{4}-12c_{6})+36c_{7}+18c_{8}\right)-12c_{6}qmr^{2}}{2\sqrt{3}r^{8}}+{\cal O}(c_i^2)\ .\nn
\eea
Corresponding to the modified solution above, the mass-charge relation of the extremal black hole becomes
\be
m_{\rm ext}=2q-4 c_4-8 c_5-8 c_6-48 c_7-24 c_8-\frac{c_1}{3}-\frac{11 c_2}{3}-\frac{31 c_3}{3}+{\cal O}(c_i^2)\ .
\label{D5mext}
\ee
The action of the test particle is given by \eqref{D4V0q} and the position-independence of the potential leads to $m=2q,\ m_{0}=\sqrt{3}q_{0}$ and additional constraints on the higher derivative couplings
\begin{align}
        & c_{4}=\frac{1}{108}(-17c_{1}+29c_{2}+97c_{3}-120c_{6}-144c_{7}-72c_{8})\ ,
\nonumber \\
        & c_{5}=\frac{1}{27}(c_{1}-16c_{2}-47c_{3}-12c_{6}-144c_{7}-72c_{8})\ .
\label{5q2}
\end{align}
 Similar to $D=4$ case, constraints from force-free condition is stronger than requiring the mass-charge relation of an extremal black hole be uncorrected, as a linear combination of the two constraints above already yields $m_{\rm ext}=2q$. Thus black holes obeying the force-free condition are still extremal. Notice that we are considering the probe limit, then these constraints are the necessary conditions for the existence of the multi-centered black hole. 
 In the next subsection, we will explore the consequences of force-free condition using the probe limit in a 3-black hole system.
    
 \subsubsection{Probe limit in the 3-black hole system}
 in order to study the probe limit in the 3-black hole system and its consequences through the force-free condition,  we again need to solve for the coefficients $ \{\delta{a}_{(i,n)},\delta{b}_{(i,n)},\delta{c}_{(i,n)},\delta{u}_{(i,n)} \}$ in $D=5$. Utilizing the solution associated with $Y_0(\cos \theta)$ (given in appendix B),  the force-free condition \eqref{V0n0} leads to the following constraints 
\begin{equation}
        c_{3}=0\ ,\quad c_{4}=\frac{1}{6}(c_{2}-c_{1})\ ,\quad c_{5}=-c_{2}\ ,\quad c_{6}=0\ ,\quad  c_{7}=\frac{1}{144}\left(c_{1}+11c_{2}-72c_{8}\right)\ .
\label{D5qConstraints}
\end{equation}
The set of constraints above is stronger than the one derived from the probe limit in the 2-black hole system. Interestingly, when these constraints are satisfied, the effective energy-momentum tensor $\Delta T_{\mu\nu}$ and effective electric current $\Delta J_\mu$ vanish not only on the 2-centered MP solution but actually on the all the uncorrected multi-centered MP solutions \eqref{TwoBH}. Therefore the MP multi-centered black hole solutions remain exact in the force-free 4-derivative extensions of Einstein-Maxwell theory. At this stage, there is no need to consider systems with more extremal black holes. 

Plugging \eqref{D5qConstraints} in \eqref{L4even}, the force-free 4-derivative couplings can be written explicitly in the form
 \begin{align}
    \Delta\mathcal{L}_{\text{force-free}}=&c_{1}(R^{2}-\frac{1}{6}RF_{\mu\nu}F^{\mu\nu}+\frac{1}{144}F_{\mu\nu}F^{\mu\nu}F_{\rho\sigma}F^{\rho\sigma}) 
\nn\\
    &+c_{2}(R_{\mu\nu}R^{\mu\nu}+\frac{1}{6}RF_{\mu\nu}F^{\mu\nu}-R_{\mu\nu}F^{\mu\rho}F^{\nu}{}_{\rho}-\frac{11}{144}({\rm tr }F^2)^2+\frac{11}{36}{\rm tr}F^4)
\nn\\
	&-\frac{1}{2}c'_{8}(F_{\mu\nu}F^{\mu\nu}F_{\rho\sigma}F^{\rho\sigma}-2F_{\mu\nu}F^{\nu\rho}F_{\rho\sigma}F^{\sigma\mu})+c_{9}\nabla^{\mu}F_{\mu\nu}\nabla_{\rho}F^{\rho\nu}\ ,
\label{L5q}
 \end{align}
where $c'_8=c_8-\frac{11}{36}c_2$.
Among the $D=5$ force-free 4-derivative combinations given above, those with coefficient $c_1$ and $c_2$ coincide precisely with the parity even terms in Ricci scalar squared \cite{Ozkan:2013nwa, Butter:2014xxa} and Ricci tensor squared  \cite{Gold:2023ymc} supergravity actions upon eliminating the auxiliary fields.
We notice that the supergravity action involving Riemann tensor squared \cite{Ozkan:2013uk,Ozkan:2013nwa} is absent in the force-free combination. The reason is that the multi-center black hole solution is $\ft12$-BPS in the 2-derivative theory and should remain so when supersymmetric higher derivative interactions are turned on.  The curvature squared supergravity invariants were initially constructed using the off-shell formulation of $D=5$ supergravity in which the BPS equation for static black holes  takes the form \cite{Ozkan:2013nwa}
\be
\r \sqrt{-g_{tt}}=A_t\ ,
\ee
where $\r$ is the scalar field inside an off-shell vector multiplet. To go to on-shell supergravity models described by $g_{\m\n}$ and $A_\m$, one has to eliminate $\r$ by field equations. 
Switching on the Weyl tensor squared supergravity action, the field equations imply that the scalar field $\rho$ is no longer a constant but depends on the field strength of $A_\m$. Thus the BPS equation $\r \sqrt{-g_{tt}}=A_t$ does not coincide with the force-free condition $\sqrt{-g_{tt}}=A_t$. Whereas for the Ricci scalar squared and Ricci tensor squared supergravity actions, $\rho=1$ still holds at the first order in the 4-derivative couplings, thus the BPS equation does imply the force-free condition. We postpone a detailed discussion on this point to Appendix C. 


    \subsubsection{Force-free constraints from magnetic strings}
 In $D=5$,  
a string can couples to the U(1) gauge field magnetically. Considering the probe limit in a 2-magnetic string system, we find that the force-free condition
leads to 4 constraints, which combined with those from the probe limit of a 2-electric black hole system, give rise to the same amount of constraints as the probe limit in a 3-black hole system \eqref{D5qConstraints}. 
To proceed, 
we first explain the details of solving for the 4-derivative corrections to a single magnetic string solution.

It is well-known that the Einstein-Maxwell theory in $D=5$ admits a magnetic black string solution ( see \cite{Noumi:2022ybv} for instance). Its generalization to the 4-derivative
theory is straightforward by using the ansatz
\begin{align}
        ds^{2}= & -f_{+}(r)dt^{2}+f_{-}(r)dx^{2}+\frac{dr^{2}}{g(r)}+r^{2}(d\theta^{2}+\sin^{2}\theta d\phi^{2})\ ,\nn
\label{D5B1BMetricAnsatz}\\
        F_{(2)}= & dA_{(1)},\quad A_{(1)}=A_{\phi}\,d\phi=-\cos(\theta)\sqrt{3r_{+}r_{-}}\,d\phi\ ,
\nonumber \\
        H_{(3)}= &dB_{(2)}=\star \widetilde{M}_{(2)},\quad B_{(2)}=B_{tx}\,dt\wedge dx, \quad \widetilde{M}^{\m\n}=-2   \frac{\partial\mathcal{L}}{\partial F_{\mu\nu}}\ ,
\end{align}
where $H_{(3)}$ is the hodge dual of $\widetilde{M}_{(2)}$. 
Substituting the ansatz (\ref{D5B1BMetricAnsatz}) into the field equations with 4-derivative corrections given in Appendix A, we find that the expression of $g(r)$
involves two logarithmic functions divergent on the horizon
\begin{align}
        g(r) & =g_{+}\text{ln}(r-r_{+})/r+g_{-}\text{ln}(r-r_{-})/r+\text{regular terms}\ ,
\label{gr}
\end{align}
where the two constants $g_{\pm}$ are given by
\bea
        g_{+} & =&\frac{1}{30r_{\text{+}}^{3}(r_{-}-r_{\text{+}})}\Big(18r_{-}^{2}(2c_{7}+c_{8})(2r_{-}-3r_{\text{+}})
\nonumber \\
         && +\frac{1}{4}\left(-(3c_{1}+33c_{2}+205c_{3})r_{-}^{2}r_{\text{+}}+140c_{3}r_{-}r_{\text{+}}^{2}+2(c_{1}+11c_{2}+45c_{3})r_{-}^{3}-80c_{3}r_{\text{+}}^{3}\right)
\nonumber \\
         && +3\left((-3c_{4}-6c_{5}+2c_{6})r_{-}^{2}r_{\text{+}}-10c_{6}r_{-}r_{\text{+}}^{2}+2(c_{4}+2c_{5}+c_{6})r_{-}^{3}-10cr_{\text{+}}^{3}\right)\Big)\ ,
\nonumber \\
            g_{-} & =&g_{+}|_{r_{+}\rightarrow r_{-}}\ ,
\eea
in which $c$ is an integration constant. 
To remove the logarithmic
divergences on the outer horizon, we set $g_{+}=0$,
yielding the constraints
\begin{equation}
       c_1= -11 c_2-59 c_3-12 \left(c_4+2 c_5+12 c_7+6 c_8\right)\ ,\quad c_{6}=\frac{7}{6}c_{3}\ ,\quad c=-\frac{2}{3}c_{3}\ .
\label{c1c6}
\end{equation}
On these constraints, $g_-$ also vanishes by virtue of the symmetry between $g_-$ and $g_+$. Imposing \eqref{c1c6} we obtain the magnetic black string solution with 4-derivative corrections 
\bea
         g(r)&= & \frac{(r-r_{-})(r-r_{+})}{r^{2}}+\frac{1}{9r^{6}}\Big(-216(2c_{7}+c_{8})r_{-}r_{+}\left(8r^{2}-7(r_{-}+r_{+})r+6r_{-}r_{+}\right)
\nonumber \\
         && +\frac{1}{2}\left(27c_{2}r_{-}r_{+}(-22r^{2}+19rr_{-}+19rr_{+}-16r_{-}r_{+})+2c_{3}(6r^{4}+3r^{3}r_{-}+3r^{3}r_{+})\right)
\nonumber \\
         &&-\frac{9}{2}r_{-}r_{+}\left(4c_{4}(8r^{2}-7rr_{-}-7rr_{+}+6r_{-}r_{+})+c_{5}(130r^{2}-113(r_{-}+r_{+})r+96r_{-}r_{+})\right)
\nonumber \\
        && +c_{3}\left(-2r^{2}(5r_{-}^{2}+821r_{+}r_{-}+5r_{+}^{2})+1401r_{-}r_{+}(r_{-}+r_{+})r-1176r_{-}^{2}r_{+}^{2}\right)\Big)+\mathcal{O}(c_i^2)\ ,\nn
\\
        f_{+}(r)&= & 1-\frac{r_{+}}{r}+\frac{1}{18r^{5}}\Big(2c_{3}r\big(12(\frac{2r}{r_{-}-r_{+}}+1)r^{2}+5r_{+}r+6r_{+}^{2}\big)
\nonumber \\
        && +16c_{3}r^{2}r_{-}-27r_{+}^{2}r_{-}\big(7c_{2}+38c_{3}+4c_{4}+15c_{5}+48(\text{2\ensuremath{c_{7}}+\ensuremath{c_{8}})}\big)
\nonumber \\
        && +4rr_{+}r_{-}\big(27c_{2}+145c_{3}+18c_{4}+63c_{5}+216(2c_{7}+c_{8})\big)\Big)+\mathcal{O}(c_i^2)\ ,\nn
\nonumber\\
        f_{-}(r)&= & 1-\frac{r_{-}}{r}+\frac{1}{9r^{5}}\Big(4c_{3}r^{2}\big(2r_{+}+3r(\frac{2r}{r_{+}-r_{-}}+1)\big)
\nonumber \\
        && +5r_{-}c_{3}r^{2}+rr_{-}r_{+}\big(54c_{2}+290c_{3}+18(2c_{4}+7c_{5})+432(2c_{7}+c_{8})\big)
\nonumber \\
        && +6c_{3}r_{-}^{2}r-\frac{27}{2}r_{-}^{2}r_{+}\big(7c_{2}+38c_{3}+4c_{4}+15c_{5}+48(2c_{7}+c_{8})\big)\Big)+\mathcal{O}(c_i^2)\ ,
\nonumber\\
      B_{ty}&=&-\frac{\sqrt{3r_{-}r_{+}}}{r}+\frac{c_{3}\sqrt{r_{-}r_{+}}\left(22r(r_{-}+r_{+})-237r_{-}r_{+}+r^{2}\right)}{3\sqrt{3}r^{5}}
\nn\\
	&&-\frac{\sqrt{3}(r_{-}r_{+})^{\frac{3}{2}}(9c_{2}+4c_{4}+17c_{5}+96c_{7}+48c_{8})}{2r^{5}}+\mathcal{O}(c_i^2)\ .
\eea
The action of a massive magnetic string takes the form
\begin{equation}
        S_{0}=-\t_{0}\int d^{2}\sigma\sqrt{-\text{det}(h_{ab})}+\frac{p_{0}}2\int B_{\mu\nu}dx^{\mu}\land dx^{\nu}\ ,
\end{equation}
where $\sigma^{a} (a=1,2)$ are the world-sheet coordinates on the string and $h_{ab}$ is the induced metric. To simplify the discussion, we choose the probe string to be parallel to the background black string solution, {\it i.e.},  $\s^0=t,\, \s^1=y$. 
Thus the force-free condition for a static probe string becomes
\begin{equation}
        V_{0}=\t_{0}\sqrt{f_{+}(r)f_{-}(r)}-p_{0}B_{tx}=\text{const}+\mathcal{O}(c_i^2)\ ,
\label{D5pV0}
\end{equation}
Substituting the modified solution to (\ref{D5pV0}), we obtain the
force-free constraints on the mass charge relation of the background string and the probe string
\begin{align}
        \t_{0}=\sqrt{3}p_{0}\ ,\quad  r_{+} & =r_{-}\ ,
\label{D5mp}
\end{align}
as well as two constraints on the higher derivative couplings
\begin{equation}
       c_{3}=0,\quad c_{4}=\frac{1}{2}\left(-3c_{2}-7c_{5}-48c_{7}-24c_{8}\right)\ ,
\label{cd5m}
\end{equation}
Combining \eqref{c1c6} and (\ref{cd5m}) we obtain the full set of constraints implied by the probe limit of 2 magnetic strings
\begin{equation}
      c_{3}=0,\ c_{4}=\frac{1}{36}\big(-7c_{1}-5c_{2}+72(2c_{7}+c_{8})\big),\ c_{5}=\frac{1}{18}\big(c_{1}-7c_{2}-72(2c_{7}+c_{8})\big),\ c_{6}=0\ .
\label{d5p}
\end{equation}
Miraculously, combining the constraints obtained above with those \eqref{5q2} from the probe limit of a 2-electric black hole system, we obtain the full set of constraints \eqref{D5qConstraints} obtained from the probe limit in a 3-electric black hole system.

\section{Conclusions and discussions}
In this work, we obtained the 4-derivative extensions of  Einstein-Maxwell theory that admit multi-extremal black hole solutions.  We did not follow the brute-force way by first solving for the 4-derivative corrections to the multi-black hole solutions and imposing certain smoothness conditions.  Instead, we adopted a more physical approach by first considering the probe limit in a 2-black hole system where one of the black hole was much lighter than the other and so could be treated as a test particle probing the background source by the heavier one. Balance between gravity and Coulomb force felt by the test particle leads to two constraints on the 4-derivative couplings, a combination of which implies that the extreme mass-charge relation is not modified by the constrained 4-derivative couplings. Taking into account backreactions from the test particle, we obtain sufficient constraints so that the resulting theory definitely admit
multi-extremal black hole solutions. We verified that force-free combinations of 4-derivative interactions do not modify the MP solution as well as the non-extremal RN black holes. However, the force-free 4-derivative do modify the dyonic black hole solutions whose thermodynamic quantities can be easily computed 
using the recent results \cite{Reall:2019sah,Ma:2023qqj}.

Since Einstein-Maxwell theory also admits magnetically charged objects, out of curiosity, we also studied consequences of force-free condition using the probe limit of 2 magnetic black holes although this seems to be unrelated to our original pursuit about existence of multi-electrically charged black holes. To our surprise, the force-free condition imposed on the magnetic black holes yields 2 constraints which together with the 2 constraints obtained from the probe limit of 2 electric black holes, leads to the same set of constraints derived from the probe limit of the 3-black hole case. Currently, we do not know if there is a deep reason behind and would like to understand it better in future.

In $D=4$, we noticed that the force-free combinations of the 4-derivative couplings include the GB term, off-shell ungauged $N=2$ supersymmetric Weyl tensor squared and the $F^4$ term appearing in quasi-topological electromagnetism. In $D=5$, the force-free combinations recover the parity even terms in two 4-derivative supergravity invariants and also the  quasi-topological $F^4$ term. 
 We recall that the Born-Infeld Lagrangian takes the form \cite{Tseytlin:1999dj}
\be
{\cal L}=-\ft14 F_{\m\n}F^{\m\n}+\ft18 \left[{\tr}(F^4)-\ft14({\tr} F^2)^2\right]+\cdots\ ,
\ee
which does not correspond to any choice of the coefficients in \eqref{D4qLeven}. Thus for Born-Infeld theory to admit multi-extremal black hole solutions, proper curvature couplings allowed by \eqref{D4qLeven} must be added.  In our approach, the field equations are solved up to first order in the 4-derivative couplings $c_i$ and we required the force-condition be satisfied at the same order. From the viewpoint of effective field theory, one can perform field redefinition without affecting the physics. Thus 4-derivative couplings related to the force-free combinations obtained here by field redefinition also allow for multi-extremal black hole solutions. The difference resides in the fact that in the redefined theory, the test particle action is no longer minimally coupled. In fact, applying field redefinitions, we find that in $D=4$ the force-free combinations reduce to Gauss-Bonnet term and the quasi-topological $F^4$ term while in $D=5$ the only irreducible structure is the quasi-topological $F^4$ term (see Appendix D). Moreover, the field redefinition employed here turns out to be an identity map on the solution space. Thus a solution in EM theory extended by the quasi-topological $F^4$ term is also a solution of \eqref{D4qLeven} and \eqref{L5q}. For the quasi-topological $F^4$ action, it is straightforward to show that it does not affect purely electric or magnetic solutions, but it does modify the dyonically charged solutions \cite{Liu:2019rib}.

As possible future research directions, we would like to generalize our analysis to other interesting models admitting multi-extremal black holes or black branes. For instance, in $D=4$, there exists the famous Garfinkle-Strominger-Horowitz multi-centered dilatonic black hole solutions \cite{Garfinkle:1990qj}. In $D=6$ the bosonic sector of $N=(1,0)$ supergravity consisting of $(g_{\m\n},\phi,\,B_{\m\n})$ admits multi-extremal string solutions \cite{Dabholkar:1990yf,Duff:1993ye}. In $D=4$, we see that for Einstein-Maxwell theory, the force-free combinations of the 4-derivative couplings include the 4-derivative supergravity invariants. In \cite{Ma:2021opb}, it was noticed that the known 4-derivative supergravity invariants \cite{Bergshoeff:1986vy, Novak:2017wqc, Butter:2018wss, Ozkan:2013cab} in $D=6$ are compatible with the force-free requirement.  The results  \cite{Ortin:2021win} from the low energy effective theory of heterotic string suggest that when a scalar field is present, the muti-centered extremal black holes or branes may be modified by the higher derivative terms and structures like Riemann tensor squared and Riemann tensor coupled to $p$-form field strength could appear in the force-free 4-derivative couplings. It should be interesting to see if the force-free condition leads to just these supergravity invariants or more structures.

\section*{Acknowledgement} 
We are grateful to discussions with Dr. Liang Ma and comunications with Dr. Alejandro Ruiperez. We also thank Prof. Hong Lu for his comments on an earlier version of this draft. This work is supported by National Natural Science Foundation of China (NSFC) under grant No. 12175164. This work is also partially supported by Peng Huanwu Center for Fundamental Theory, under grant No.12247103. Y.P. also acknowledges support by the National Key Research and Development Program under grant No. 2022YFE0134300.

\appendix
\section{Corrections to equations of motion}

In this work, we considered the following action
\begin{equation}
       S_{{\rm EM}+4\partial}=\int d^{D}x\sqrt{-g}\frac{1}{2\kappa^{2}}(R-\frac{1}{4g^{2}}F_{\mu\nu}F^{\mu\nu}+ \Delta\mathcal{L})\ ,
\end{equation}
and $\Delta\mathcal{L}$ is
\begin{align}
        \Delta\mathcal{L}= & c_{1}R^{2}+c_{2}R_{\mu\nu}R^{\mu\nu}+c_{3}R_{\mu\nu\rho\sigma}R^{\mu\nu\rho\sigma}\;
\nn\\
        + & c_{4}RF_{\mu\nu}F^{\mu\nu}+c_{5}R_{\mu\nu}F^{\mu\rho}F^{\nu}{}_{\rho}+c_{6}R_{\mu\nu\rho\sigma}F^{\mu\nu}F^{\rho\sigma}\;
\nonumber \\
        + & c_{7}F_{\mu\nu}F^{\mu\nu}F_{\rho\sigma}F^{\rho\sigma}+c_{8}F_{\mu\nu}F^{\nu\rho}F_{\rho\sigma}F^{\sigma\mu}+c_{9}\nabla^{\mu}F_{\mu\nu}\nabla_{\rho}F^{\rho\nu}\ ,
\nn \\
        +&d_{4}RF_{\mu\nu}\widetilde{F}^{\mu\nu}+d_{5}R_{\mu\nu}F^{\mu\rho}\widetilde{F}_{\ \rho}^{\nu}+d_{6}R_{\mu\nu\rho\sigma}F^{\mu\nu}\widetilde{F}^{\rho\sigma}
\nn    \\                       +&d_{7}\widetilde{F}_{\mu\nu}F^{\mu\nu}F^{\rho\sigma}F_{\rho\sigma}+d_{8}\widetilde{F}_{\mu\nu}F^{\nu\rho}F_{\rho\sigma}F^{\sigma\mu}\ .
\end{align}
 where $F=dA,\ \widetilde{F}=\star F$. In the field equations given below, the term with coefficient $c_{9}$
is not included, because $\nabla^\mu F_{\m\n}$ always vanishes  on the leading order solution. The field equations are given by
\begin{align}
        g_{\mu\nu}: & \ R_{\mu\nu}-\frac{1}{2}Rg_{\mu\nu}=\frac{1}{2g^{2}}(T_{\mu\nu}+ \Delta T_{\mu\nu})\ ,
\label{Einodd}\\
        A_{\mu}: & \ \nabla_{\mu}F^{\mu\nu}=\Delta J^\nu\ ,
\label{Maxwellodd}
\end{align}
 where $T_{\mu\nu}=F_{\mu}^{\ \rho}F_{\nu\rho}-\frac{1}{4}g_{\mu\nu}F_{\rho\sigma}F^{\rho\sigma}$. The electric current $\Delta J^\nu=\nabla_{\mu} M^{\mu\nu}$ with  $M^{\mu\nu}$ given by
\begin{align}
        M^{\mu\nu} & =2\frac{\partial\Delta\mathcal{L}}{\partial F_{\mu\nu}}
\nonumber \\
       & =4(c_{4}RF^{\mu\nu}-c_{5}R_{\ \rho}^{[\mu}F^{\nu]\rho}+c_{6}R^{\mu\nu\rho\sigma}F_{\rho\sigma}+2c_{7}F^{\mu\nu}F^{\rho\sigma}F_{\rho\sigma}+2c_{8}F^{\mu\rho}F^{\nu\sigma}F_{\rho\sigma})\ , 
\nn\\
        &+4d_{4}R\widetilde{F}^{\mu\nu}+d_{5}(R^{\rho[\mu}\widetilde{F}_{\rho}^{\ \nu]}+R_{\ \sigma}^{\rho}F_{\rho\tau}\epsilon^{\sigma\tau\mu\nu})
\nn\\
        & +2d_{6}(R^{\mu\nu\rho\sigma}\widetilde{F}_{\rho\sigma}+\frac{1}{2}\epsilon^{\mu\nu\lambda\tau}R_{\lambda\tau\rho\sigma}F^{\rho\sigma})+4d_{7}(F^{\mu\nu}\widetilde{F}_{\rho\sigma}\widetilde{F}^{\rho\sigma}+\widetilde{F}^{\mu\nu}F_{\rho\sigma}F^{\rho\sigma})
\nn\\
        & +2d_{8}(\frac{1}{2}\epsilon^{\mu\nu\lambda\tau}F_{\rho\lambda}F_{\sigma\tau}F^{\rho\sigma}+\widetilde{F}_{\rho\sigma}F^{\mu\rho}F^{\nu\sigma}+2F_{\ \ \rho}^{[\mu}\widetilde{F}^{\nu]\sigma}F_{\ \sigma}^{\rho})\ .
\label{Muv}     
\end{align}
The effective energy-momentum tensor  $\Delta T_{\mu\nu}$ is 
\begin{align}
        \Delta T_{\mu\nu}  &=\frac{-2}{\sqrt{-g}}\frac{\delta(\sqrt{-g}\Delta{\cal L})}{\delta g^{\mu\nu}}= -2\big(P_{(\mu}^{\ \alpha\beta\gamma}R_{\nu)\alpha\beta\gamma}-2\nabla^{\rho}\nabla^{\sigma}P_{\rho(\mu\nu)\sigma}-\frac{1}{2}\Delta{\cal L}g_{\mu\nu}+\frac{1}{2}M_{(\mu}^{\ \alpha}F_{\nu)\alpha}\big)
\nonumber \\
        P^{\mu\nu\rho\sigma}& =\frac{\partial\Delta{\cal L}}{\partial R_{\mu\nu\rho\sigma}} 
\nn\\        
        &=2c_{1}Rg^{\mu[\rho}g^{\sigma]v}+c_{2}(R^{\mu[\rho}g^{\sigma]\nu}-R^{\nu[\rho}g^{\sigma]\mu})+2c_{3}R^{\mu\nu\rho\sigma}
\nonumber \\
        & +c_{4}g^{\mu[\rho}g^{\sigma]v}F_{\alpha\beta}F^{\alpha\beta}+c_{5}\frac{1}{2}(S^{\mu[\rho}g^{\sigma]\nu}-S^{\nu[\rho}g^{\sigma]\mu})+c_{6}F^{\mu\nu}F^{\rho\sigma}
\nn\\
        &+d_{4}g^{\mu[\rho}g^{\sigma]v}F_{\alpha\beta}\widetilde{F}^{\alpha\beta}+d_{5}\frac{1}{2}(\widetilde{S}^{\mu[\rho}g^{\sigma]\nu}-\widetilde{S}^{\nu[\rho}g^{\sigma]\mu})+d_{6}\frac{1}{2}(F^{\mu\nu}\widetilde{F}^{\rho\sigma}+F^{\rho\sigma}\widetilde{F}^{\mu\nu})  \ ,   
        \label{Tuv}
\end{align}
where $S^{\mu\nu}=F_{\ \rho}^{\mu}F^{\nu\rho}, \widetilde{S}^{\mu\nu}=F_{\ \ \rho}^{(\mu}\widetilde{F}^{\nu)\rho}$. In the derivation, we find the following formulae are useful
\bea
          \frac{\partial R_{\alpha\beta\gamma\delta}}{\partial R_{\mu\nu\rho\sigma}}&=&\frac{1}{2}\left(\delta_{[\alpha}^{\mu}\delta_{\beta]}^{\nu}\delta_{[\gamma}^{\rho}\delta_{\delta]}^{\sigma}+\delta_{[\gamma}^{\mu}\delta_{\delta]}^{\nu}\delta_{[\alpha}^{\rho}\delta_{\beta]}^{\sigma}\right)\ ,
\nn\\
  \frac{\partial R^{2}}{\partial R_{\mu\nu\rho\sigma}}&=&2Rg^{\alpha\gamma}g^{\beta\delta}\frac{\partial R_{\alpha\beta\gamma\delta}}{\partial R_{\mu\nu\rho\sigma}}=2Rg^{\mu[\rho}g^{\sigma]v}\ ,
\nn\\
          \frac{\partial(R_{\alpha\beta}R^{\alpha\beta})}{\partial R_{\mu\nu\rho\sigma}}&=&2R^{\alpha\beta}g^{\gamma\delta}\frac{\partial R_{\alpha\beta\gamma\delta}}{\partial R_{\mu\nu\rho\sigma}}=R^{\mu[\rho}g^{\sigma]\nu}-R^{\nu[\rho}g^{\sigma]\mu}\ ,
\nn\\
          \frac{\partial(R_{\alpha\beta\gamma\delta}R^{\alpha\beta\gamma\delta})}{\partial R_{\mu\nu\rho\sigma}}&=&2R^{\alpha\beta\gamma\delta}\frac{\partial R_{\alpha\beta\gamma\delta}}{\partial R_{\mu\nu\rho\sigma}}=2R^{\mu\nu\rho\sigma}\ .
\eea

\section{$\delta a_{(i,0)},\ \delta b_{(i,0)},\ \delta c_{(i,0)},\ \delta u_{(i,0)} $}
In $D=4$, up to ${\cal O}(r^{-5})$ the nonzero coefficients in the spherical harmonic expansion at $n=0$ are
\begin{align}
        \delta a_{(0,0)} & =\frac{64}{75m_{1}^{2}}\big(-155c_{2}-620c_{3}+1676c_{4}+330c_{5}+728c_{6}+528\left(2c_{7}+c_{8}\right)\big)
\nonumber \\
        & +\frac{64m_{2}}{1875m_{1}^{3}}\Big(20065c_{2}+80260c_{3}-4\big(51667c_{4}+10185c_{5}+24226c_{6}+18276\left(2c_{7}+c_{8}\right)\big)\Big)\ ,
\nn\\
        \delta a_{(2,0)} & =\frac{4m_{2}}{625m_{1}}\Big(1165c_{2}+4660c_{3}-16588c_{4}-16\big(240c_{5}+479c_{6}+429\left(2c_{7}+c_{8}\right)\big)\Big)
\nonumber \\
        & +\frac{4}{25}\big(-5c_{2}-20c_{3}+116c_{4}+30c_{5}+48\left(c_{6}+2c_{7}+c_{8}\right)\big)\ ,
\nn\\
        \delta a_{(3,0)} & =\frac{-4m_{1}}{75}\left(10c_{2}+40c_{3}+8c_{4}+15c_{5}-c_{6}+48c_{7}+24c_{8}\right)
\nonumber \\
         & +\frac{-4m_{2}}{1875}\big(895c_{2}+3580c_{3}-15544c_{4}-3795c_{5}-6832c_{6}-6432\left(2c_{7}+c_{8}\right)\big)\ ,
\nn\\
        \delta a_{(4,0)} & =\frac{1}{200}m_{1}^{2}\big(85c_{2}+340c_{3}-2\left(266c_{4}+30c_{5}+173c_{6}+96c_{7}+48c_{8}\right)\big)
\nonumber \\
        & +\frac{1}{1250}m_{1}m_{2}\big(605c_{2}+2420c_{3}-13256c_{4}-3330c_{5}-6443c_{6}-4968\left(2c_{7}+c_{8}\right)\big)\ ,
\nn
\end{align}

\begin{align}
        \delta a_{(5,0)} & =\frac{1}{1600}m_{1}^{3}\big(-205c_{2}-820c_{3}+2256c_{4}+455c_{5}+2168c_{6}+768\left(2c_{7}+c_{8}\right)\big)
\nonumber \\
         & +\frac{m_{1}^{2}m_{2}}{40000}\big(1265c_{2}+5060c_{3}+142192c_{4}+44685c_{5}+135176c_{6}+62976\left(2c_{7}+c_{8}\right)\big),\\
    \delta{c}_{(0,0)} & =\frac{16}{375m_{1}^{2}}\Big(7940c_{2}+31760c_{3}-8\big(10546c_{4}+2055c_{5}+4588c_{6}+3288\left(2c_{7}+c_{8}\right)\big)\Big)
\nonumber \\
        &  +\frac{48m_{2}}{9375m_{1}^{3}}\Big(16\big(-21985c_{3}+47523c_{4}+9265c_{5}+28094c_{6}+23044\left(2c_{7}+c_{8}\right)\big)-87940c_{2}\Big)\ ,
\nn\\
         \delta{c}_{(1,0)} & =\frac{256}{125m_{1}}\Big(115c_{2}+460c_{3}-1228c_{4}-6\big(40c_{5}+89c_{6}+64\left(2c_{7}+c_{8}\right)\big)\Big)
\nonumber \\
         & -\frac{4096m_2}{9375m_{1}^{2}}\Big(760c_{2}+3040c_{3}-6772c_{4}-1335c_{5}-4\big(979c_{6}+804\left(2c_{7}+c_{8}\right)\big)\Big)\ ,
\nn \\
\delta{c}_{(2,0)} & =\frac{4m_{2}}{9375m_{1}}\big(-190955c_{2}-763820c_{3}+1790276c_{4}+358680c_{5}+994628c_{6}+816528\left(2c_{7}+c_{8}\right)\big)
\nonumber \\
         & +\frac{4}{375}\Big(4735c_{2}+2\big(9470c_{3}-25646c_{4}-5055c_{5}-11138c_{6}-8088\left(2c_{7}+c_{8}\right)\big)\Big)\ ,
\nn\\
\delta{c}_{(3,0)} & =\frac{4}{25}m_{1}\big(20c_{2}+80c_{3}-224c_{4}-45c_{5}-97c_{6}-72\left(2c_{7}+c_{8}\right)\big)
\nonumber \\
         & +\frac{12}{625}m_{2}\big(-395c_{2}-1580c_{3}+3944c_{4}+795c_{5}+2032c_{6}+1632\left(2c_{7}+c_{8}\right)\big)\ ,
\nn\\
         \delta{c}_{(4,0)} & =\frac {m_{1}} {16}\left(c_{2}+4c_{3}+8c_{4}+3c_{5}+4c_{6}\right)\left(m_{1}+2m_{2}\right)\ ,
\nn\\
         \delta{c}_{(5,0)} & =-\frac{m_{1}^{2}}{320}\left(9c_{2}+36c_{3}+80c_{4}+29c_{5}+40c_{6}\right)\left(m_{1}+3m_{2}\right)\ ,
\\
  \delta{u}_{(1,0)} &=\frac{4}{3m_{1}}\Big(26c_{2}+104c_{3}-4\left(74c_{4}+15c_{5}+8\big(4c_{6}+6c_{7}+3c_{8}\right)\big)\Big)
\nonumber \\
	& +\frac{16m_{2}}{15m_{1}^{2}}\big(-145c_{2}-580c_{3}+1552c_{4}+315c_{5}+736c_{6}+576\left(2c_{7}+c_{8}\right)\big)\ \nn,
\\
	\delta{u}_{(2,0)} & =\frac{m_{2}}{625m_{1}}\Big(2330c_{2}+9320c_{3}-33176c_{4}-32\big(240c_{5}+479c_{6}+429\left(2c_{7}+c_{8}\right)\big)\Big)
\nonumber \\
	& +\frac{1}{25}\big(-10c_{2}-40c_{3}+232c_{4}+60c_{5}+96\left(c_{6}+2c_{7}+c_{8}\right)\big)\ ,
\nn\\
	\delta{u}_{(3,0)} & =\frac{1}{150}\left(-55c_{2}-220c_{3}+316c_{4}+30c_{5}+148c_{6}+96c_{7}+48c_{8}\right)m_{1}
\nonumber \\
	& +\frac{m_{2}}{1875}\big(-230c_{2}-920c_{3}+10556c_{4}+2955c_{5}+3968c_{6}+4368\left(2c_{7}+c_{8}\right)\big)\ ,
\nn\\
        \delta{u}_{(4,0)}&=\ft{m_1}{3000} \big(-50\left(-50c_{3}+98c_{4}+15c_{5}+59c_{6}+48c_{7}+24c_{8}\right)m_{1}+5c_{2}\left(125m_{1}+37m_{2}\right)\big)
\nonumber \\
	& -\frac{1}{750}m_{1}m_{2}\big(-185c_{3}+1283c_{4}+315c_{5}+749c_{6}+324\left(2c_{7}+c_{8}\right)\big)\ ,
\nn\\
	\delta{u}_{(5,0)} & =\frac{m_{1}^{3}}{3200}\big(-245c_{2}-980c_{3}+2192c_{4}+415c_{5}+2056c_{6}+896\left(2c_{7}+c_{8}\right)\big)
\nonumber \\
	&-\frac{m_{1}^{2}m_{2}}{80000}\big(5395c_{2}+21580c_{3}-84544c_{4}-21045c_{5}-109832c_{6}-42432\left(2c_{7}+c_{8}\right)\big), 
\end{align}
where $\delta{b}_{(i,0)}=\delta{c}_{(i,0)}$ and  $\delta{u}_{(0,0)}$  can't be determined from EOM due to the U(1) gauge symmetry. In $D=5$, up to ${\cal O}(r^{-18})$ the nonzero coefficients in the spherical harmonic expansion at $n=0$ are
\begin{align}
\text{\ensuremath{\delta}}a_{(0,0)}= & 2\text{\ensuremath{\delta}c}_{(0,0)}+\frac{4}{245m_{1}}\left(1753c_{1}+18092c_{2}+56207c_{3}+41196c_{4}+81201c_{5}\right)\nonumber \\
 & +\frac{288}{245m_{1}}\big(6791c_{6}+5113(2c_{7}+c_{8})\big)\:,\nn\\
\text{\ensuremath{\delta}}a_{(2,0)}= & \frac{1}{5880}(-180945c_{1}+1075146c_{2}+3453529c_{3}-30348c_{4}+3004845c_{5})\nonumber \\
 & +\frac{1}{5880}\big(16369032c_{6}+12663864(2c_{7}+c_{8})+5145m_{1}\text{\ensuremath{\delta}c}_{(0,0)}\big)\nonumber \\
 & -\frac{3m_{2}}{245m_{1}}\big(39170c_{3}+30600c_{4}+55261c_{5}+327536c_{6}+245424(2c_{7}+c_{8})\big)\nonumber \\
 & -\frac{m_{2}}{490m_{1}}\left(10148c_{1}+75994c_{2}+245m_{1}\text{\ensuremath{\delta}c}_{(0,0)}\right)\ ,\nn\\
 \text{\ensuremath{\delta}}a_{(4,0)}= & \frac{3m_{2}}{23520}\big(1019108c_{4}+768501c_{5}+4902504c_{6}+3585240(2c_{7}+c_{8})\big)\nonumber \\
 & +\frac{m_{2}}{23520}\left(360169c_{1}+152714c_{2}+338543c_{3}-2205m_{1}\text{\ensuremath{\delta}c}_{(0,0)}\right)\nonumber \\
 & -\frac{m_{1}}{23520}\left(-238393c_{1}+759214c_{2}+2481697c_{3}-854124c_{4}+1673289c_{5}\right)\nonumber \\
 & -\frac{m_{1}}{23520}\big(8875080c_{6}+6914808(2c_{7}+c_{8})+2205m_{1}\text{\ensuremath{\delta}c}_{(0,0)}\big)\ \nn,\\
 \text{\ensuremath{\delta}}a_{(6,0)}= & \frac{3m_{1}^{2}}{94080}\big(524331c_{3}-500452c_{4}+208949c_{5}+1135832c_{6}+902952(2c_{7}+c_{8})\big)\nonumber \\
 & +\frac{1}{94080}(-735m_{1}^{3}\text{\ensuremath{\delta}c}_{(0,0)}-287849m_{1}^{2}c_{1}+463220m_{1}^{2}c_{2})\nonumber \\
 & -\frac{m_{1}m_{2}}{6720}\left(48317c_{1}-16544c_{2}-71445c_{3}+336540c_{4}+125049c_{5}\right)\nonumber \\
 & -\frac{m_{1}m_{2}}{6720}\big(778584c_{6}+559656\left(2c_{7}+c_{8}\right)-315m_{1}\text{\ensuremath{\delta}c}_{(0,0)}\big)\ ,\nn\\
 \text{\ensuremath{\delta}}a_{(8,0)}= & \frac{3m_{1}^{3}}{376320}\big(594628c_{4}+52667c_{5}+73960c_{6}-15336\left(2c_{7}+c_{8}\right)\big)\nonumber \\
 & +\frac{m_{1}^{3}}{376320}\left(3675m_{1}\text{\ensuremath{\delta}c}_{(0,0)}+297953c_{1}-132284c_{2}-539257c_{3}\right)\nonumber \\
 & +\frac{9m_{1}^{2}m_{2}}{376320}\big(674644c_{4}+169951c_{5}+738440c_{6}+475512\left(2c_{7}+c_{8}\right)\big)\nonumber \\
 & +\frac{3m_{1}^{2}m_{2}}{376320}\left(317509c_{1}-45412c_{2}-273661c_{3}-1225m_{1}\text{\ensuremath{\delta}c}_{(0,0)}\right)\ ,\nn\\
\text{\ensuremath{\delta}}a_{(10,0)}= & \frac{-3m_{1}^{4}}{1505280}\big(477076c_{4}+240805c_{5}+772936c_{6}+381432\left(2c_{7}+c_{8}\right)\big)\nonumber \\
 & -\frac{m_{1}^{4}}{1505280}\left(6615m_{1}\text{\ensuremath{\delta}c}_{(0,0)}+222645c_{1}+309054c_{2}+885091c_{3}\right)\nonumber \\
 & -\frac{20m_{1}^{3}m_{2}}{1505280}\left(44307c_{1}+42276c_{2}+115685c_{3}+248100c_{4}\right)\nonumber \\
 & -\frac{20m_{1}^{3}m_{2}}{1505280}\Big(51099c_{5}-24\big(4889c_{6}+8871\left(2c_{7}+c_{8}\right)\big)\Big)\ \nn ,
\end{align}

\begin{align}
 \text{\ensuremath{\delta}}a_{(12,0)}= & \frac{m_{1}^{5}}{6021120}\left(1361c_{1}+956638c_{2}+3042855c_{3}+92940c_{4}+1127547c_{5}\right)\nonumber \\
 & +\frac{m_{1}^{5}}{6021120}\big(3311352c_{6}+1017288\left(2c_{7}+c_{8}\right)+9555m_{1}\text{\ensuremath{\delta}c}_{(0,0)}\big)\nonumber \\
 & +\frac{3m_{1}^{4}m_{2}}{6021120}\big(-758540c_{4}-42531c_{5}-6178296c_{6}-7139784\left(2c_{7}+c_{8}\right)\big)\nonumber \\
 & +\frac{m_{1}^{4}m_{2}}{6021120}\left(11289675c_{3}+6615m_{1}\text{\ensuremath{\delta}c}_{(0,0)}-81779c_{1}+3524078c_{2}\right)\ ,\nn\\
 \text{\ensuremath{\delta}}a_{(14,0)}= & \frac{3m_{1}^{6}}{24084480}\big(886740c_{4}-485239c_{5}-1250744c_{6}+60024\left(2c_{7}+c_{8}\right)\big)\nonumber \\
 & +\frac{m_{1}^{6}}{24084480}\left(-12495m_{1}\text{\ensuremath{\delta}c}_{(0,0)}+440869c_{1}-1926598c_{2}-6353965c_{3}\right)\nonumber \\
 & +\frac{6m_{1}^{5}m_{2}}{24084480}\left(473163c_{1}-1535826c_{2}-5137795c_{3}+3528900c_{4}+317181c_{5}\right)\nonumber \\
 & +\frac{6m_{1}^{5}m_{2}}{24084480}\big(6985896c_{6}+8279064\left(2c_{7}+c_{8}\right)-2695m_{1}\text{\ensuremath{\delta}c}_{(0,0)}\big)\ \nn,\\
 \text{\ensuremath{\delta}}a_{(16,0)}= & \frac{3m_{1}^{7}}{96337920}\big(-2445204c_{4}+603695c_{5}+1440056c_{6}-701688\left(2c_{7}+c_{8}\right)\big)\nonumber \\
 & +\frac{m_{1}^{7}}{96337920}\left(11315421c_{3}+15435m_{1}\ensuremath{\delta}c_{(0,0)}-1193365c_{1}+3355294c_{2}\right)\nonumber \\
 &+ \frac{m_{1}^{6}m_{2}}{96337920}\left(-8673535c_{1}+19839802c_{2}+67864023c_{3}-59527476c_{4}-3806265c_{5}\right)\nonumber \\
&+ \frac{m_{1}^{6}m_{2}}{96337920}\big(-69392136c_{6}-89835192\left(2c_{7}+c_{8}\right)+28665m_{1}\ensuremath{\delta}c_{(0,0)}\big)\ ,\nn\\
\text{\ensuremath{\delta}}a_{(18,0)}= & \frac{m_{1}^{8}}{1156055040}\left(7742746c_{1}-15485383c_{2}-53485954c_{3}+47565048c_{4}\right)-2\text{\ensuremath{\delta}c}_{(18,0)}\nonumber \\
 & +\frac{m_{1}^{8}}{1156055040}\big(-5525493c_{5}-15368880c_{6}+13302864\left(2c_{7}+c_{8}\right)-55125m_{1}\text{\ensuremath{\delta}c}_{(0,0)}\nonumber \\
 & +\frac{m_{1}^{7}m_{2}}{289013760}\left(15835916c_{1}-27095528c_{2}-94922924c_{3}-33075m_{1}\text{\ensuremath{\delta}c}_{(0,0)}\right)\nonumber \\
 & +\frac{48m_{1}^{7}m_{2}}{24084480}\big(8654924c_{4}+535631c_{5}+6232600c_{6}+8843592\left(2c_{7}+c_{8}\right)\big)\ ,\\
 \text{\ensuremath{\delta}c}_{(2,0)}= & \frac{-3}{11760}\left(649020c_{4}+2300831c_{5}+13279576c_{6}+10111464\left(2c_{7}+c_{8}\right)+2695m_{1}\text{\ensuremath{\delta}c}_{(0,0)}\right)\nonumber \\
 & +\frac{m_{2}}{10m_{1}}\left(32c_{1}+37c_{2}+104c_{3}+192c_{4}+69c_{5}+96c_{6}\right)\nonumber \\
 & +\frac{1}{11760}(96801c_{1}-1943562c_{2}-6151465c_{3})\ ,\nn\\
\text{\ensuremath{\delta}c}_{(4,0)}= & \frac{-3m_{2}}{47040}\big(381484c_{4}+1846101c_{5}+10579320c_{6}+8124552\left(2c_{7}+c_{8}\right)\big)\nonumber \\
 & +\frac{m_{2}}{47040}\left(-11025m_{1}\text{\ensuremath{\delta}c}_{(0,0)}+147781c_{1}-1623808c_{2}-5166109c_{3}\right)\nonumber \\
 & -\frac{3m_{1}}{47040}\big(621532c_{4}+2197953c_{5}+12572760c_{6}+9597096\left(2c_{7}+c_{8}\right)\big)\nonumber \\
 & -\frac{m_{1}}{47040}\left(11025m_{1}\text{\ensuremath{\delta}c}_{(0,0)}-89113c_{1}+1884424c_{2}+5962897c_{3}\right)\ ,\nn
\end{align}

\begin{align}
\text{\ensuremath{\delta}c}_{(6,0)}= & \frac{1}{24}\left(4c_{1}+5c_{2}+14c_{3}+24c_{4}+9c_{5}+12c_{6}\right)m_{1}\left(m_{1}+2m_{2}\right)\ ,\nn\\
\text{\ensuremath{\delta}c}_{(8,0)}= & -\frac{1}{384}\left(32c_{1}+37c_{2}+104c_{3}+192c_{4}+69c_{5}+96c_{6}\right)m_{1}^{2}\left(m_{1}+3m_{2}\right)\ \nn,\\
\text{\ensuremath{\delta}c}_{(10,0)}= & \frac{m_{1}^{3}\left(m_{1}+4m_{2}\right)}{1280}\left(241c_{4}+86c_{5}+122c_{6}+12c_{7}+6c_{8}\right)\nonumber \\
 & \frac{m_{1}^{3}\left(m_{1}+4m_{2}\right)}{15360}\big(481c_{1}+539c_{2}+1519c_{3}\big)\ ,\nn\\
\text{\ensuremath{\delta}c}_{(12,0)}= & -\frac{m_{1}^{4}\left(m_{1}+5m_{2}\right)}{24576}\big(257c_{1}+283c_{2}+799c_{3}+1548c_{4}+552c_{5}\big)\nonumber \\
 & -\frac{3m_{1}^{4}\left(m_{1}+5m_{2}\right)}{1024}\big(11c_{6}+2c_{7}+c_{8}\big)\ ,\nn\\
\text{\ensuremath{\delta}c}_{(14,0)}= & \frac{m_{1}^{5}\left(m_{1}+6m_{2}\right)}{14336}\big(283c_{4}+101c_{5}+146c_{6}+36c_{7}+18c_{8}\big)\nonumber \\
 & +\frac{m_{1}^{5}\left(m_{1}+6m_{2}\right)}{172032}\big(563c_{1}+613c_{2}+1733c_{3}\big)\ ,\nn\\
\text{\ensuremath{\delta}c}_{(16,0)}= & -\frac{m_{1}^{6}\left(m_{1}+7m_{2}\right)}{65536}\big(389c_{4}+139c_{5}+202c_{6}+60c_{7}+30c_{8}\big)\nonumber \\
 & -\frac{m_{1}^{6}\left(m_{1}+7m_{2}\right)}{786432}\big(773c_{1}+835c_{2}+2363c_{3}\big)\ \\
 \text{\ensuremath{\delta}}u_{(2,0)}= & \frac{m_{2}}{10m_{1}}\left(32c_{1}+37c_{2}+104c_{3}+192c_{4}+69c_{5}+96c_{6}\right)\nonumber \\
 & -\frac{1}{11760}(-39353c_{1}+2259494c_{2}+7123297c_{3}+2770836c_{4}+8234049c_{5})\nonumber \\
 & -\frac{1}{11760}\big(47332680c_{6}+36083448\left(2c_{7}+c_{8}\right)+13965m_{1}\text{\ensuremath{\delta}c}_{(0,0)}\big)\ ,\nn\\
 \text{\ensuremath{\delta}}u_{(4,0)}= & \frac{3m_{1}}{11760}\big(80986c_{3}+68648c_{4}+110963c_{5}+624496c_{6}+479088\left(2c_{7}+c_{8}\right)\big)\nonumber \\
 & +\frac{3m_{2}}{11760}\big(80986c_{3}+68648c_{4}+110963c_{5}+624496c_{6}+479088\left(2c_{7}+c_{8}\right)\big)\nonumber \\
 &- \frac{1}{11760}\big(-735m_{1}^{2}\text{\ensuremath{\delta}c}_{(0,0)}-14362c_{1}\left(m_{1}+m_{2}\right)-78983c_{2}\left(m_{1}+m_{2}\right)\big)\ ,\nn\\
\text{\ensuremath{\delta}}u_{(6,0)}= & -\frac{m_{1}^{2}}{31360}\left(10856c_{1}+42799c_{2}+130544c_{3}+123552c_{4}+170487c_{5}+926944c_{6}\right)\nonumber \\
 & -\frac{m_{1}m_{2}}{15680}\left(9103c_{1}+24707c_{2}+74337c_{3}+82356c_{4}+89286c_{5}+437992c_{6}\right)\nonumber \\
 & -\frac{m_{1}}{15680}\big((350496m_{1}+332856m_{2})\left(2c_{7}+c_{8}\right)+245m_{1}^{2}\text{\ensuremath{\delta}c}_{(0,0)}\big)\ ,\nn\\
 \text{\ensuremath{\delta}}u_{(8,0)}= & \frac{m_{1}^{3}}{94080}\left(9593c_{1}+27157c_{2}+81687c_{3}+88236c_{4}+98106c_{5}+512472c_{6}\right)\nonumber \\
 & -\frac{m_{1}^{3}}{188160}\big(-736272\left(2c_{7}+c_{8}\right)-735m_{1}\text{\ensuremath{\delta}c}_{(0,0)}\big)\nonumber \\
 & +\frac{m_{1}^{2}m_{2}}{128}\left(32c_{1}+37c_{2}+104c_{3}+192c_{4}+69c_{5}+96c_{6}\right)\ \nn,
\end{align}

\begin{align}
 \text{\ensuremath{\delta}}u_{(10,0)}= & -\frac{3m_{1}^{4}}{1505280}\big(127424c_{4}+113669c_{5}+565088c_{6}+374784\left(2c_{7}+c_{8}\right)+490m_{1}\text{\ensuremath{\delta}c}_{(0,0)}\big)\nonumber \\
 & -\frac{m_1^3}{1505280}\Big( m_{1}\left(48096c_{1}+105519c_{2}+312824c_{3}\right)-240m_{2}\big(8417c_{6}+8871\left(2c_{7}+c_{8}\right)\big)\Big)\nonumber \\
 & -\frac{m_{1}^{3}m_{2}}{150528}\left(16083c_{1}+9642c_{2}+23957c_{3}+78756c_{4}-9759c_{5}\right)\ ,\nn\\
\text{\ensuremath{\delta}}u_{(12,0)}= & \frac{49m_{1}^{5}}{1003520}\left(217c_{1}+428c_{2}+1263c_{3}+1644c_{4}+1329c_{5}+6328c_{6}\right)\nonumber \\
 & -\frac{m_{1}^{4}}{1003520}\Big(8m_{2}\big(50681c_{6}+58383\left(2c_{7}+c_{8}\right)\big)-49m_{1}\big(4104\left(2c_{7}+c_{8}\right)+5m_{1}\text{\ensuremath{\delta}c}_{(0,0)}\big) \Big)\nonumber \\
 & +\frac{m_{1}^{4}m_{2}}{1003520}\left(46153c_{1}+32492c_{2}+84607c_{3}+237996c_{4}+801c_{5}\right)\ ,\nn\\
\text{\ensuremath{\delta}}u_{(14,0)}= & -\frac{m_{1}^{6}}{1605632}\big(45428c_{4}+38379c_{5}+179336c_{6}+124728\left(2c_{7}+c_{8}\right)+98m_{1}\text{\ensuremath{\delta}c}_{(0,0)}\big)\nonumber \\
 & -\frac{m_{1}^{5}}{4816896}\big(m_{1}\left(17517c_{1}+35256c_{2}+104731c_{3}\right)+m_{2}168\left(563c_{1}+613c_{2}+1733\right)\big)\nonumber \\
 & -\frac{3m_{1}^{5}m_{2}}{7168}\left(283c_{4}+101c_{5}+146c_{6}+36c_{7}+18c_{8}\right)\ ,\nn\\
\text{\ensuremath{\delta}}u_{(16,0)}= & +\frac{m_{1}^{7}}{8028160}\big(65934c_{3}+83512c_{4}+77527c_{5}+360784c_{6}+272592\left(2c_{7}+c_{8}\right)\big)\nonumber \\
 & +\frac{m_{1}^{7}}{48168960}\left(60796c_{1}+131774c_{2}+735m_{1}\text{\ensuremath{\delta}c}_{(0,0)}\right)\nonumber \\
  & +\frac{2m_{1}^{6}m_{2}}{48168960}\big(3175896c_{6}+2411208\left(2c_{7}+c_{8}\right)\big)\nonumber \\
 & +\frac{2m_{1}^{6}m_{2}}{48168960}\left(197009c_{1}+298381c_{2}+878751c_{3}+1382988c_{4}+897258c_{5}\right)\ .
\end{align}
 where $\delta{b}_{(i,0)}=\delta{c}_{(i,0)}$. Although $\delta{c}_{(0,0)},\ \delta{u}_{(0,0)}$ cannot be determined by EOM at $\mathcal{{O}}({r^{-18}})$, the force-free condition requires $\delta{c}_{(0,0)}=\delta{u}_{(0,0)}=0$.
\section{Weyl tensor squared action in $D=5$ supergravity }
In the section, we show that the scalar field $\rho$ that appears in $D=5$, $N=2$ supergravity cannot be set to constant by field equations.  We begin our discussions with the 2-derivative action in $D=5$, $N=2$ supergravity. After eliminating several auxiliary fields, one ends up with the action below \cite{Ozkan:2013nwa}[eq.8.3]
\bea
e^{-1}{\cal L}_{2\partial}&=&\ft18(\r^3+3)R+\ft13(104\r^3-8)T_{\m\n}T^{\m\n}+4(\r^3-1)D+\ft34\r F_{\m\n}F^{\m\n}
\nn\\
&&+\ft32\r\partial^\m \r\partial_\m \r-12\r^2F_{\m\n}T^{\m\n}+\ft18\epsilon^{\m\n\r\s\l}A_\m F_{\n\r}F_{\s\l}\ .
\eea
where comparing to \cite{Ozkan:2013nwa}[eq.8.3], we have set $I=J=K=1$, $C_{111}=1$ to focus on the minimal case. The field equations of $D$, $T_{ab}$ and $\r$ imply
\be
\r=1,\quad D=0,\quad T_{\m\n}=\ft3{16}F_{\m\n}\ .
\label{2dsol}
\ee
 Substituting the solutions above back to action above, we obtain the action of minimal supergravity in $D=5$
\be
e^{-1}{\cal L}_{2\partial}=\ft12 R-\ft38 F_{\m\n}F^{\m\n}+\ft18\epsilon^{\m\n\r\s\l}A_\m F_{\n\r}F_{\s\l}\ .
\ee
To recover the familar form, one needs to rescale $F_{ab}\rightarrow \frac1{\sqrt 3}F_{ab}$. The relevant terms in Weyl squared action are
\be
e^{-1}{\cal L}_{Weyl}=\left(\ft18 \r C_{\m\n\r\l}C^{\m\n\r\l}+\ft{64}3\r D^2+\ft{1024}9\r T_{\m\n}T^{\m\n}D-\ft{32}3 D T_{\m\n}F^{\m\n}+\cdots\right)\ ,
\ee
where we have ignored terms that are independent of the field $D$. Now we consider 
the model ${\cal L}_{2\partial}+a_1{\cal L}_{Weyl}$ where $a_1$ is a small coefficient. The $D$-equation then implies 
\be
0=4(\r^3-1)+a_1 (\ft{128}3\r D+\ft{1024}9\r T_{\m\n}T^{\m\n}-\ft{32}3T_{\m\n}F^{\m\n})\ .
\ee
 Up to first order in $\a_1$, one finds 
 \be
 \r=1+\ft{a_1}2 F^{(0)}_{\m\n}F^{(0)\m\n}\ ,
 \ee
 for any $0$-th order solution of $F_{\m\n}$.

On the other hand, the relevant terms in the Ricci scalar squared action are \cite{Ozkan:2013nwa}
\be
e^{-1}{\cal L}_{R^2}=\r \left(\ft9{64}R^2-3DR+16D^2-2RT_{ab}T^{ab}+\ft{64}3DT_{\m\n}T^{\m\n} +\ft{64}9 (T_{\m\n}T^{\m\n})^2+\cdots\right)\ ,
\ee
whose correction to $D$-equation is proportional to $R-\ft14 F_{\m\n}F^{\m\n}$ which vanishes on the leading order field equations. Thus inclusion of the Ricci scalar squared invariant will not affect the value of $\r$. Also upon substituting \eqref{2dsol} into ${\cal L}_{R^2}$ and performing the rescaling
$F_{ab}\rightarrow \frac1{\sqrt 3}F_{ab}$, we recover the first combination satisfying the force-free condition in five dimensions. The analysis for the Ricci tensor squared invariant \cite{Gold:2023ymc} is very similar and  will not be repeated here. 

Further evidence can be seen from the explicit multicenter solution given in \cite{Ortin:2021win}. Focusing on the $D=5$ case, the solution is of the form 
\bea
ds_E^2&=&-(Z_+Z_-Z_0)^{-2/3}dt^2+(Z_+Z_-Z_0)^{1/3}dx^mdx^m\ ,
\nn\\
A^{(1)}&=&Z_0^{-1}dt,\quad A^{(2)}=Z^{-1}_+dt,\quad A^{(3)}=Z^{-1}_-dt\ ,
\nn\\
e^{-2\phi}&=&Z_+^{1/2}Z_-^{1/2}Z^{-1}_0,\quad k=Z_+^{1/2}Z_-^{-1/2}\ ,
\eea
where we have set the unnecessary constants $\phi_{\infty}=0$ and $k_{\infty}=1$ for simplicity and dualized the 2-form $B_{\m\n}$ to the 1-form $A^1$ as already suggested in \cite{Ortin:2021win}. \cite{Ortin:2021win} noticed that  when the particular supersymmetric
4-derivative interactions involving Riemann tensor squared are turned on, the ansatz above remains the same, only the $Z$-functions receive corrections. The ansatz above implies 3 force-free or BPS equations
\be
e^{-4/3\phi}\sqrt{-g_{tt}}=A^{(1)}_t,\quad e^{2/3\phi}k^{-1}\sqrt{-g_{tt}}=A^{(2)}_t,\quad  e^{2/3\phi}k\sqrt{-g_{tt}}=A^{(3)}_t\ .
\ee
In the 2-derivative theory, the field equations allow the solution with 
$Z_+=Z_-=Z_0$. In this case, we see that the scalar fields indeed become constant
and the force-free condition is simply $\sqrt{-g_{tt}}=A_t$. However, when 4-derivatives
interactions are switched on, the filed equations no longer allow us to set the scalar fields to constant values. Even if we choose $Z_+=Z_-=Z_0=Z^{(0)}$ at the 2-derivative level, the 4-derivative interactions modify the solution to
\cite{Ortin:2021win}, 
\begin{align}
{Z}_{+} & ={Z}^{(0)}-\frac{\alpha^{'}}{2}\delta{Z}+\mathcal{O}(\alpha'^{2}),\quad 
\delta{Z}=\frac{\partial_{m}{Z}^{(0)}\partial_{m}{Z}^{(0)}}{{Z}^{(0)2}}
\nonumber \\
{Z}_{0} & = {Z}^{(0)}+\frac{\alpha^{'}}{2}\delta{Z}+\mathcal{O}(\alpha'^{2}),\quad
{Z}_{-}={Z}^{(0)}+\mathcal{O}(\alpha'^{2})\ ,
\nonumber \\
e^{-2\phi} & =({Z}_{+}{Z}_{-})^{\frac{1}{2}}{Z}_{0}^{-1}=(1-\alpha'\frac{3\delta{Z}}{4{Z}^{(0)}})+\mathcal{O}(\alpha'^{2})\ ,
\nonumber \\
k & =(\frac{{Z}_{+}}{{Z}_{-}})^{\frac{1}{2}}=(1-\alpha'\frac{\delta{Z}}{4{Z}^{(0)}})+\mathcal{O}(\alpha'^{2})\ .
\label{SolutionSupergravity}
\end{align}
Now, in our notation, 
\begin{equation}
H=({Z}_{+}{Z}_{-}{Z}_{0})^{\frac{1}{3}}={Z}^{(0)}+\mathcal{O}(\alpha'^{2})\ ,\label{H}
\end{equation}
which means the metric in MP solution in 
the $D=5$ supergravity is uncorrected up to first order in $\alpha'$. However,
the scalar fields $e^{-2\phi},k$ are not constant and will play important roles in the force-free condition. 
\section{Irreducible force-free combinations from field redefinitions}

In $D=4$, after performing the field redefinition, 
\bea
g_{\m\n} &\rightarrow& g_{\m\n}+\delta g_{\m\n},\quad A_\m\rightarrow A_\m+\delta A_\m\ ,
\nn\\
\delta g_{\m\n}&=& (c_2+4c_3)R_{\m\n}-\ft12(2c_1+c_2+2c_3)g_{\m\n}R
\nn\\
&&-\ft12(c_2+4c_3)F_{\m\r}F_{\n}{}^\r+\ft18(c_2+4c_3)g_{\m\n}F^{\a\b}F_{\a\b}\ ,
\nn\\
&=&(c_2+4c_3)\left(R_{\m\n}-\ft 12 g_{\m\n}R-\ft 12 T_{\m\n}  \right)-(c_1-c_3) g_{\m\n}R\ ,
\nn\\
\delta A_{\n}&=&-c_9 \nabla^\m F_{\m\n}\ ,
\eea
the 4-derivative Lagrangian \eqref{D4qLeven} becomes
\be
\Delta {\cal L}=c_3 \mathcal{L}_{\rm GB} +\ft{1}{8} (c_2+4 c_3-4 c_8) \left[({\tr} F^2)^2-2{\tr}(F^4)\right]\ ,
\label{4dqt}
\ee 
where $\mathcal{L}_{\rm GB}$ denotes the standard Gauss-Bonnet term which is topological.  The quartic $F$ term corresponds to the combination of the quasi-topological electromagnetism. In $D=4$, we recall that 
\be
(\star F_{\m\n}F^{\m\n})^2=4{\tr}(F^4)-2({\tr} F^2)^2\ ,
\ee
whose contributions to the field equations automatically vanishes for all purely electric or magnetic solutions. Since GB term does not affect the field equations, it is clear that the multi-centered static charged black holes in the 2-derivative theory remains exact solutions when the 4-derivative interactions take the form in \eqref{4dqt}. Moreover, it can be checked 
that $\delta g_{\m\n}=0,\, \delta A_\m=0$ after substituting the leading order EOM, which means the solution is not modified by the field redefinitions above. Thus it is evident that up to first order in $c_i$, the multi-centered electric black hole solution remains uncorrected up to first order in $c_i$ in Einstein-Maxwell theory extended by \eqref{D4qLeven}. 

In $D=5$, after performing the field redefinition, 
\bea
g_{\m\n} &\rightarrow& g_{\m\n}+\delta g_{\m\n},\quad A_\m\rightarrow A_\m+\delta A_\m\ ,
\nn\\
\delta g_{\m\n}&=& c_2R_{\m\n}-\ft13(2c_1+c_2)g_{\m\n}R
\nn\\
&&-\ft12c_2F_{\m\r}F_{\n}{}^\r+\ft1{18}(c_1+2c_2)g_{\m\n}F^{\a\b}F_{\a\b}\ ,
\nn\\
\delta A_{\n}&=&c_{9}(-\nabla^{\mu}F_{\mu\nu}+\frac{3}{4}\mu_{\text{CS}}\epsilon_{\ \ \ \ \ \nu}^{\rho\sigma\lambda\tau}F_{\rho\sigma}F_{\lambda\tau})\ ,
\label{5dfred}
\eea
the 4-derivative Lagrangian \eqref{L5q} becomes
\be
\Delta {\cal L}=\ft{1}{8} (c_2-4 c_8) \left[({\tr} F^2)^2-2{\tr}(F^4)\right]\ .
\label{5dqt}
\ee
Similar to $D=4$, the $F^4$ above does not affect purely electric or magnetic solutions.
Also, $\delta g_{\m\n}=0,\,\delta A_\m=0$ when the leading order field equations are satisfied.

\bibliographystyle{utphys}
\bibliography{ref}

\end{document}